\documentclass{rspublic}
\usepackage{amsmath}
\usepackage{amsfonts}
\usepackage{amssymb}
\usepackage[Symbol]{upgreek}
\usepackage[nointegrals]{wasysym}
\usepackage{mathrsfs}
\usepackage{graphicx}
\newcommand{\real}{{\mathbb R}}
\newcommand{\bfx}{{\bf x}}
\newcommand{\bfy}{{\bf y}}
\newcommand{\bfz}{{\bf z}}
\newcommand{\bfu}{{\bf u}}
\newcommand{\bfV}{{\bf V}}
\newcommand{\bfB}{{\bf B}}
\newcommand{\bfC}{{\bf C}}
\newcommand{\bfF}{{\bf F}}
\newcommand{\bfG}{{\bf G}}
\newcommand{\bfK}{{\bf K}}
\newcommand{\bfH}{{\bf H}}
\newcommand{\bfQ}{{\bf Q}}
\newcommand{\bfR}{{\bf R}}
\newcommand{\starsp}{\:\star\:}
\begin{document}
\title[Navier-Stokes Equations in Cylindrical Co-ordinates]{\large Vorticity evolution in a rigid pipe of circular cross-section}
\author[F. Lam]{F. Lam}
\label{firstpage}
\maketitle
\vspace{2mm}
\begin{quote}
{\it Yet every exact solution of the equations of fluid mechanics can actually occur in Nature. The flows must not only obey the equations which are non-linear, but also evolve in the distinct dynamic states as observed}.
\end{quote}
\vspace{3mm}
\begin{abstract}{Navier-Stokes Equations; Vorticity; Pipe Hagen-Poiseuille Flow; Laminar Flow; Transition; Turbulence; Reynolds Number; Randomness}

In this paper, we show that the spatio-temporal evolution of incompressible flows in a long circular pipe can be described by vorticity dynamics. The principal techniques to obtain solutions are similar to those used for flows in $\real^3$. As the consideration of the Navier-Stokes equations is given in a cylindrical co-ordinates system, two aspects of complication arise. One is the interaction of the velocity components in the radial and azimuthal directions, due to the fictitious centrifugal force in the equations of motion. The rate of the vorticity production at the pipe wall depends on the initial data at entry, and hence is unknown {\it a priori}; it must be determined as part of the solution. The vorticity solution obtained defines an intricate flow-field of multitudinous degrees of freedom. As the Reynolds number increases, the analytical solution predicts vorticity-scale proliferations in succession. For sufficiently large initial data, pipe flows are of a turbulent nature. The solution of the governing equations is globally regular and does not bifurcate in space or in time. It is asserted that laminar-turbulent transition is a dynamic process inbred in the non-linearity. The presence of exogenous disturbances, due to imperfect test environments or purpose-made artificial forcing, distorts the course of the intrinsic transition. The flow structures observed by Reynolds (1883) and others can be synthesised and elucidated in light of the current theory.
\end{abstract}
\tableofcontents
\section{Introduction}
In the Eulerian description of the motion of an incompressible, homogeneous Newtonian fluid, the momentum and the continuity equations for fluid dynamics are 
\begin{equation} \label{ns}
	{\partial {\bf u}}/{\partial t} + ({\bf u} . \nabla) {\bf u} = \nu \Delta {\bf u}   - {\rho}^{-1} \nabla p, \;\;\; \nabla . {\bf u} =0,
\end{equation}
where the velocity vector ${\bf u}={\bf u}({\bfx},t)$ is the velocity, the scalar quantity $p=p({\bf x},t)$ is the pressure, the space variable is denoted by ${\bfx}{=}(x,y,z)$, and $\Delta$ is the Laplacian. The density and the viscosity of the fluid are denoted by $\rho$ and $\mu$ respectively. The kinematic viscosity is $\nu {=}\mu/\rho$.  These equations are known as the Navier-Stokes equations (Navier 1823; Stokes 1845). They are derived on the basis of the continuum hypothesis (see, for example, Lamb 1975; Batchelor 1967). 

We seek the solution of (\ref{ns}) as an initial-boundary value problem in the space-time domain denoted by $\Upomega {\times} [0,t>0]$, where $\Upomega$ has a smooth impermeable boundary $\partial \Upomega$. 
The initial condition is given by
\begin{equation} \label{nsic}
 \bfu(\bfx,t{=}0) = {\bfu}_0({\bfx})\; \in C^{\infty}_c\;\;\;\;{\bfx} \in \Upomega,
\end{equation} 
and the no-slip boundary condition is
\begin{equation} \label{nsbc}
 \bfu(\bfx,t) = 0\;\;\; {\bfx} \in \partial \Upomega.
\end{equation} 
The postulation that $\bfu_0$ is a smooth function with compact support is to simplify the technicalities in the subsequent analysis.

The vector quantity, known as the vorticity ${\omega}=(\omega_1,\omega_2,\omega_3)$, is related to the velocity by ${\omega}= \nabla {\times} {\bf u}$. The vorticity field must be solenoidal because of the vector formula $\nabla.(\nabla {\times} \bfu)=0$. The vorticity is a direct consequence of conservation of the angular momentum of fluid particles. Because of fluid viscosity, particles are subjected to shearing force during motion. Locally, they behave like a rigid body in rotation. At the same time, their moments of inertia redistribute as a result of deformation. The two effects give rise to the dynamic equations, 
\begin{equation} \label{nsvort}
	{\partial \omega_i} / {\partial t} - \nu \Delta \omega_i = (\omega.\nabla)\bfu_i -(\bfu.\nabla)\omega_i,\;\;\;i=1,2,3.
\end{equation}
Because the pressure has been eliminated, the vorticity equations are more accessible in deriving {\it a priori} bounds, which are critical in global regularity theory of the Navier-Stokes equations (Lam 2013). In particular, it is verified that the total vorticity is an invariant in any fluid motion with smooth initial velocity of bounded energy,
\begin{equation*}
\frac{\rd}{\rd t} \int_{\real^3} \big(\omega_1+ \omega_2+ \omega_3 \big) \rd \bfx = 0.
\end{equation*}
Consequently every vorticity component must be finite. By virtue of the Sobolev embedding theorem, the invariant principle can be extended to show that
\begin{equation} \label{vspacereg}
\omega (\bfx,t) \in C_B^{\infty}(\bfx)\; C_B^{\infty}(t),\;\;\; t \geq t_c > 0,
\end{equation}
where $t_c$ is the small time limit for smoothness in the classical regularity theory. Since the vorticity is a fundamental quantity in fluid motion, we need an inverse relation which enables us to compute the velocity and the pressure. In view of the incompressibility hypothesis and of the vector identity 
\begin{equation} \label{vid}
\nabla \times \omega=\nabla(\nabla.{\bfu}) - \Delta {\bfu},
\end{equation}
an elliptic equation can be derived and used to bound $\nabla \bfu$ in space because di-vorticity, $\nabla{\times}\omega$, is a sub-set of Jacobian, $\nabla \omega$. The space-regularity in the velocity gradients ensures the pressure gradient $\nabla p$ is suitably bounded. From the momentum equation, we assert that
\begin{equation} \label{vtimereg}
	\big\|\nabla \bfu(\cdot,t)\big\|_{L^{\infty}(\real^3)}  \leq C \: (\nu t)^{-3/4},\;\;\;t>0.
\end{equation}
In conjunction with the classical local-in-time regularity, the global well-posedness follows for initial smooth data of finite energy (see, for example, Doering \& Gibbon 1995; Constantin 2001; Heywood 2007). By a similarity reduction, it has been shown that, at high Reynolds number, the vorticity field is nothing more than an amalgamation of {\it viscous} shearing motions, giving rise to multitudinous spatio-temporal scales. In the sequel, we shall work with infinitely differentiable bounded functions, as no finite-time singularity can develop in solution of the Navier-Stokes equations. Viscosity of fluids, no matter how small it may be, activates energy consumption; the non-linearity in the equations facilitates the dissipation by self-multiplying viscous shears into smaller sizes. In fact, the continuity defines a zero-sum of three {\it extended reals}, $\partial u_i/\partial x_i,i{=}1,2,3$ at any given instant of time. None of them can go out of bounds as the ``measure" $\infty{-}\infty{=}a$, where $a$ is a {\it finite} real, is undefined in analysis. The physics is clear; any motion due to an incompressible flow of finite energy cannot instigate an unbounded plenum adjacent to an infinity vacuum, in any finite region whose flow rate of matter is finite. 

The space-time vorticity boundedness, $\omega \in L^{\infty}(\bfx)\; L^{\infty}(t)$, holds in the limit of vanishing viscosity $\nu {\rightarrow} 0$. Hence the Euler equations ($\mu{=}0$) cannot blow up in finite time according to the BKM theorem (Beale {\it et al.} 1984), assuming that there exists a fluid of zero viscosity. The impossibility of singularities in inviscid flows of finite energy is consistent with Helmholtz's vortex theorems (1858) and Kelvin's circulation principle (1869). Owing to lack of energy dissipation, flow motions described by the Euler can only re-distribute their energy. Thus, it is comprehensible that, at least locally, inviscid flow fields may evolve into less smooth states compared to their viscous counter-parts. Nevertheless, the Euler equations do not constitute a self-contained model for the physics of turbulence, where diffusive dissipation by viscosity is of importance in vorticity proliferation. 
\section{Velocity-vorticity formulation}
In the present paper, we deal with flows in the presence of solid surfaces and we must strictly impose the no-slip boundary condition (\ref{nsbc}). We first establish that, in the ``kinematic" system 
\begin{equation} \label{kinem}
\nabla \times \bfu = \omega,\;\;\; \nabla.\bfu=0,\;\;\; \bfu = 0\;\; \bfx \in \partial \Upomega,
\end{equation}
the velocity can be recovered uniquely from 
\begin{equation} \label{kinem-poi}
\Delta \bfu = - \nabla \times \omega,
\end{equation}
where the region $\Upomega$ is simply-connected having a $C^2$ boundary $\partial \Upomega$.
The necessary condition to confirm (\ref{kinem-poi}) is the vector identity (\ref{vid}). The continuity implies that the vorticity field in $\Upomega$ is solenoidal $\nabla.\omega=0$. Denote $G$ by Green's function satisfying the homogeneous boundary condition. Conversely, if we are given a bounded vector function $A$ in $\Upomega$.  The Poisson equation, $\Delta \bfu = -\nabla{\times}A$, is inverted 
\begin{equation} \label{bs}
\bfu(\bfx)=\int_{\Upomega} G(\bfx,\bfx') \nabla_{\bfx'} {\times} A(\bfx') \rd \bfx' =  -\int_{\Upomega} \nabla_{\bfx}  {\times} G(\bfx',\bfx) A(\bfx') \rd \bfx',
\end{equation}
where the boundary term from integration by parts vanishes because function $G$ is zero on the boundary, and the last integral holds in view of the symmetric property of Green's function: $G(\bfx,\bfx')=G(\bfx',\bfx)$.
Thus the incompressibility $\nabla.\bfu=0$ is verified if we take divergence operation. Taking curl on the Poisson equation driven by $A$, we obtain
\begin{equation*}
\Delta (\nabla \times \bfu) = \Delta A - \nabla (\nabla.A).
\end{equation*}
If the vector $A$ is solenoidal and coincides with $\omega$ a.e., then the function $A=\omega+A_b$ generally holds such that $A\equiv\omega$, and 
$\Delta A_b=0, \; \nabla.A_b=0$, and this system has no boundary conditions. Given $A_b$ as a vector, we must have $\nabla{\times}\nabla{\times} A_b=0$ or the last elliptic system reduces to
\begin{equation*}
\nabla.A_b=0,\;\;\;\nabla \times A_b=a,
\end{equation*}
where $a$ is a constant. Thus the following condition can be established:
\begin{equation} \label{bdy-vort}
A_b(\bfx \in \partial\Upomega) = a \int_{\Upomega} \rd \bfx = a \; {\mbox{vol}}(\Upomega),
\end{equation}
which defines a vortex sheet over the boundary. In application, we have two scenarios: (1) The volume $\Upomega$ is unbounded. Since we are interested in fluid motions of finite energy, there can be no singular vortex sheets on the boundary in view of (\ref{vspacereg}) and (\ref{vtimereg}). The function $A$ is nothing but the vorticity. (2) For domains of finite $\mbox{vol}(\Upomega)$, the function $A_b$ can be fixed by initial data (say at $t{=}0$). The problem is to trace the spatio-temporal flow development from {\it a set of initial-boundary conditions} by solving the Navier-Stokes equations. In particular, if the initial vorticity in $\Upomega$ is zero, the system of (\ref{kinem}) and (\ref{kinem-poi}) is well-posed in any singly-connected region with smooth boundaries; the velocity can be recovered uniquely from the vorticity (\ref{bs}). Our ultimate aim is to solve the dynamic equation (\ref{ns}) or (\ref{nsvort}) in $\Upomega$ where 
both $\bfu$ and $\omega$ exist and evolve in {\it real fluids}. The use of the dynamic no-slip boundary condition is therefore justified, as homogeneous Dirichlet condition (\ref{nsbc}) is valid at every given instant of time.

In addition, imposing (\ref{nsbc}) has another advantage which simply fixes the velocity field in the Helmholtz decomposition. Let us write
\begin{equation*}
\bfu = \nabla \phi + \nabla \times \uppsi.
\end{equation*}
The function $\uppsi$ can be determined by taking curl operation so that $\Delta \uppsi = - \nabla \times \bfu = -\omega$ as long as the $\uppsi$-field is solenoidal. The continuity demands that $\phi$ is harmonic
\begin{equation*}
\Delta \phi = 0.
\end{equation*}
In $\real^3$, it is postulated that the ``potential" $\phi$ decays at infinity, or we consider only finite energy solutions as a requirement for physics. Thus, a harmonic function bounded in $\real^3$ must be a constant by virtue of Liouville's theorem. The contribution from the irrotational part drops out implicitly. For real fluids in $\Upomega$, the no-slip condition adds the boundary condition $\nabla \phi(\bfx)=0, \;\forall \bfx \in \partial \Upomega$. The solvability constraint for the Laplace equation,
\begin{equation*}
\int_{\partial \Upomega} \nabla \phi (\bfx) \rd \bfx =0,
\end{equation*}
holds. Hence the only solution for the equation is a constant. Briefly, the formulation (\ref{kinem})-(\ref{kinem-poi}) uniquely specifies the velocity or the vorticity field for real fluids $\mu {>} 0 $.
\section{Dynamic equations in cylindrical co-ordinates}
In a cylindrical co-ordinates system $(r, \theta, z)$, we consider the initial-boundary value of flow evolution. The velocity and pressure $(\bfu,p)({\bfx},t)=(\bfu,p)(r, \theta, z, t)$. Let $\bfu=(u,v,w)$ denote the velocity components in the radial, circumferential and axial directions respectively. Without loss of generality, we take the pipe radius (denoted by $a$) as unity. The domain of interest $\Upomega$ consists of the interior of a semi-infinite circular cylinder
\begin{equation} \label{domain}
0 \leq r \leq 1,\;\;\;0 \leq \theta < 2 \pi,\;\;\; 0 \leq z < \infty.
\end{equation}
We denote the wall at $r=1$ in $\Upomega$. The cross-section at the location $z=0$ is the pipe entry. The pipe surface is assumed to be hydraulically smooth. The choice of the co-ordinates attempts to model, as closely as possible, practical laboratory conditions. In many pipe-flow experiments, controlled water or air from a reservoir is forced into the pipe under external pressure differences. By experience, good flows may be generated in practice by maintaining a constant mass flow through the entire pipe test section. At the same time, care must be taken to minimise fluctuations in the flow rate. The flow far downstream is inevitably in turbulent stage and the presence of the turbulence needs to be accounted for in theoretical treatment. The axial length is typically $100$-$1000$ times of pipe diameter, depending on specific set-up. Although the practical pipe length may not be long enough for the flow to decay completely at the far end of the pipe, the flow field close to the inlet must be, to a good approximation, independent of downstream flow.
\subsection*{Momentum equations}
The continuity equation reads
\begin{equation} \label{incmp}
\nabla.{\bfu} = \frac{1}{r}\frac{\partial (ru)} {\partial r} + \frac{1}{r}\frac{\partial v} {\partial \theta} + \frac{\partial w} {\partial z} =0.
\end{equation} 
The Navier-Stokes momentum equations are
\begin{equation} \label{nseq}
\begin{split}
{\partial_t u}  + ({\bfu}.\nabla) u - \frac{v^2}{r} & = -\frac{1}{\rho}\frac{\partial p}{\partial r} + \nu \Big( \Delta' u - \frac{2}{r^2} \frac{\partial v}{\partial \theta} \Big),\\
{\partial_t v} + ({\bfu}.\nabla) v + \frac{u\:v}{r} & = -\frac{1}{\rho r}\frac{\partial p}{\partial \theta} + \nu \Big( \Delta' v + \frac{2}{r^2} \frac{\partial u}{\partial \theta} \Big),\\
{\partial_t w} + ({\bfu}.\nabla) w & = - \frac{1}{\rho}\frac{\partial p}{\partial z} + \nu \Delta w,
\end{split}
\end{equation}
where $\Delta $ is the cylindrical Laplacian  
\begin{equation} \label{lap}
\Delta = \frac{\partial^2 } {\partial r^2} + \frac{1}{r} \frac{\partial } {\partial r}  + \frac{1}{r^2}\frac{\partial^2 } {\partial \theta^2} + \frac{\partial^2} {\partial z^2},
\end{equation}
and the differential operator $\Delta'=\Delta {-}{1}/{r^2}$, and $\nabla B=(\partial_r B, \partial_{\theta}B /r , \partial_z B)$ for a scalar $B$ (see, for example, Batchelor 1967; Schlichting 1979; Wu {\it et al.} 2006).

The pressure $p$ satisfies the following Poisson equation:
\begin{equation} \label{ppoi}
\begin{split}
\frac{\Delta p} {\rho} = & \;\;2 \;\Big( \frac{\partial u} {\partial r}\Big)^2 + 2 \; \Big( \frac{1}{r}\frac{\partial v} {\partial \theta} {+} \frac{u}{r}\Big)^2 + 2 \; \Big( \frac{\partial w} {\partial z}\Big)^2 \\
& \;\; + \Big( \frac{1}{r}\frac{\partial u} {\partial \theta} {+} \frac{\partial v}{\partial r} {-} \frac{v}{r}\Big)^2 
+ \Big( \frac{\partial u} {\partial z} {+} \frac{\partial w}{\partial r}\Big)^2 + \Big( \frac{\partial v} {\partial z} {+} \frac{1}{r}\frac{\partial w}{\partial \theta}\Big)^2.
\end{split}
\end{equation}
Hence the pressure can be determined uniquely with respect to a reference pressure, once the velocity gradients are known. The elliptic equation contains no time-wise information; any variation in the velocity gradients causes instantaneous changes in the pressure. This apparently unphysical causality effect is a consequence of the continuity (\ref{incmp}). If the density of the fluid is assumed to vary with the pressure, as given in a state equation, the pressure variation propagates at the local speed of sound. In standard laboratory conditions, the speeds of sound for pure water and air are about $1500\:$\texttt{m/s} and $340\:$\texttt{m/s} respectively. Thus, the incompressibility is a well-suited hypothesis in pipe flow experiments. 
 
We suppose that the motion starts impulsively from rest at time $t=0$, and the flow is initiated afterwards. The initial condition is taken as
\begin{equation} \label{ic0}
{\bfu}\;(r,\theta,z>0,t \leq 0)=0.
\end{equation}
Equations (\ref{incmp}) and (\ref{nseq}) are to be solved subject to the following boundary conditions $\forall t \in (0,T< \infty]$. The velocity must satisfy the no-slip condition at wall
\begin{equation} \label{bc}
{\bfu}\;(r=1,\theta,z>0, t>0)=0,
\end{equation}
and decays at infinity
\begin{equation} \label{decay}
{\bfu} \rightarrow 0\;\;\; \mbox{as}\;\;\; z \rightarrow \infty.
\end{equation}
The decay is plausible in physics for flows of finite energy. In addition, the velocity  inside the pipe is assumed to be bounded everywhere, 
\begin{equation}\label{bd}
\big|\;{\bf u}\;(0 \leq r < 1, \theta, z)\;\big| < \infty.
\end{equation}
Because of the azimuthal symmetry, it is evident that the velocity must satisfy the periodic condition
\begin{equation} \label{bc2}
\bfu\;(r,\theta=0,z)=\bfu\;(r,\theta=2\pi,z),\;\;\; (r,z) \in \Upomega
\end{equation}
at every instant of time. For $t\geq0$, we postulate that the flow condition at the pipe entry can be specified as smooth functions of position and time
\begin{equation} \label{ic}
{\bfu}\;(r,\theta,z=0,t)={\bfu}_e\:(r,\theta,t)=(\;u_e\;\;\;v_e\;\;\;w_e\;)(r,\theta,t).
\end{equation}
Since the entry velocity is also governed by the equations of motion, we require that $\nabla.{\bfu}_e=0$, and the initial energy is finite.
\subsection*{Vorticity equations}
The vorticity components in the cylindrical co-ordinates are 
\begin{equation} \label{vort}
\xi = \frac{1}{r}\frac{\partial w}{\partial \theta} - \frac{\partial v}{\partial z},\;\;\; 
\eta = \frac{\partial u}{\partial z} - \frac{\partial w}{\partial r},\;\;\;
\zeta = \frac{1}{r}\frac{\partial (rv)}{\partial r} - \frac{1}{r}\frac{\partial u}{\partial \theta}
\end{equation}
in the $r$, $\theta$, $z$ directions respectively. In the vector identity, 
$\nabla {\times} \omega=\nabla(\nabla.{\bfu}) {-} \Delta {\bfu}$,
the quantity on the left-hand side is known as the di-vorticity which links the velocity and the vorticity 
\begin{equation} \label{v-poi}
\Delta {\bf u} = - \nabla {\times} \omega
\end{equation}
for incompressible flows. In terms of the di-vorticity, the Navier-Stokes momentum equations (\ref{nseq}) can be re-written as
\begin{equation} \label{cylns}
\partial_t{\bfu} + \omega \times {\bfu} + \nu {\nabla {\times} \omega} = \nabla \chi,
\end{equation}
where $\chi=-({\bfu}^2/2{+}p/\rho)$, is the Bernoulli-Euler pressure (cf. (\ref{ppoi})). Taking the curl on (\ref{cylns}) and (\ref{incmp}), and making use of the vector identity $\nabla.(\nabla{\times}{\bf u})=0$, we obtain the dynamics equations for the vorticity components 
\begin{equation} \label{vdyn}
\begin{split}
{\partial_t \xi} -  \nu \Delta' \xi &= ({\omega} . \nabla \big) u - ({\bfu}.\nabla) \xi - \frac{{2}\nu}{r^2} \frac{\partial \eta}{\partial \theta}= X - \frac{{2}\nu}{r^2} \frac{\partial \eta}{\partial \theta}= \bar{X},\\
{\partial_t \eta} - \nu \Delta' \eta  &= ({\omega} . \nabla \big) v  - ({\bfu}.\nabla) \eta + \frac{{2} \nu}{r^2} \frac{\partial \xi}{\partial \theta} + \frac{{\eta u} - {\xi v}}{r}=Y + \frac{{2} \nu}{r^2} \frac{\partial \xi}{\partial \theta}=\bar{Y},\\
{\partial_t \zeta}   -  \nu \Delta \zeta &= ({\omega} . \nabla \big) w-({\bfu}.\nabla) \zeta = Z=\bar{Z},
\end{split}
\end{equation}
where the vector components $(X\;\;Y\;\;Z)={\bf X}$, and $(\bar{X}\;\;\bar{Y}\;\;\bar{Z})=\bar{\bf X}$. Note that the continuity equation is satisfied. In addition, the pressure drops out of the vorticity formulation (\ref{vdyn}) so that the number of unknowns is now three, instead of four. The dynamic pressures, $\partial_t p,\;\partial_t \partial_r p$ and the other derivatives, must be found, indirectly, from one of the momentum equations in (\ref{nseq}).

As our motion starts impulsively from rest, condition (\ref{ic0}) means the initial data for vorticity are identically zero
\begin{equation} \label{vic0}
\omega\;(r,\theta,z>0,t \leq 0)=0.
\end{equation}
At any subsequent instant during the evolution, the vorticity satisfies the periodic condition
\begin{equation} \label{vbc}
\omega\;(r,\theta=0,z)=\omega\;(r,\theta=2\pi,z).
\end{equation}
From the velocity distribution (\ref{ic}), the vorticity at the entry can be calculated
\begin{equation} \label{vic}
\nabla \times {\bf u}_e\:(r,\theta,z=0,t)=\omega_e\:(r,\theta,t),
\end{equation}
where we have explicitly expressed $\bfu_e$ or $\omega_e$ as a function of time. The axial gradient of the entry vorticity is given by
\begin{equation} \label{dvic}
\frac{\partial \nabla \times {\bf u}_e}{\partial z}\Big|_{z=0}=\upsilon_e\:(r,\theta,t)=(\;f_e\;\;\;g_e\;\;\;h_e\;)(r,\theta,t).
\end{equation}

The left-hand sides in (\ref{vdyn}) are diffusion operators. During the vorticity evolution, the pipe wall acts as a source for vorticity in view of the no-slip condition. However, the amount of the vorticity generated across the wall per unit area per unit time is not known and clearly depends on the dynamics of the whole flow field. Theoretically, the wall vorticity must constitute part of the solution. For convenience, we designate the vorticity boundary value as
\begin{equation} \label{vwall}
\omega\;(r=1,\theta,z,t)=\omega_b\:(\theta,z,t).
\end{equation}

To simplify our notations, the following short-hand is helpful:
\begin{equation*}
\int_{\Upomega}=\int_0^1 \int_0^{2 \pi} \!\!\!\int_0^{\infty}\;\;,\;\;\;\;  \int_{\partial\Upomega} = \int_0^{2 \pi} \!\!\!\int_0^{\infty} \;\;,\;\;\;\;\int_{\Upomega^{\bigotimes}} = \int_0^1 \int_0^{2 \pi},
\end{equation*}
where the middle integral is over the entire pipe wall and it is used to specify the wall vorticity while the last over the inner cross-sectional area.
Similarly, space-time integrals are
\begin{equation*}
\int_{\Upomega_T}=\int_0^t\int_0^1 \int_0^{2 \pi} \!\!\!\int_0^{\infty}\;\;,\;\;\;\; \int_{\partial\Upomega_T} = \int_0^t \int_0^{2 \pi} \!\!\!\int_0^{\infty}\;\;,\;\;\;\;
\int_{\Upomega^{\bigotimes}_T}=\int_0^t\int_0^1 \int_0^{2 \pi}\;\;.
\end{equation*}
In our presentation, we use the following notations for the independent variables
\begin{equation*}
\bfx = (r,\theta,z),\;\;\;\; \bfy=(r,\theta),\;\;\;\;\bfz=(\theta,z).
\end{equation*}
Use of them becomes clear when the domain of integration is stated.

In the present paper, we intend to develop an analytical theory for the flow-field evolution from the commencement of motion, given the prescribed data. The fluid dynamics in circular pipes has been subjected to intensive investigations for more than $130$ years, starting with Reynolds' experiments (Reynolds 1883). We refer the reader to a recent review, and the references therein, on the latest development (Mullin 2011). It is an experimental fact that pipe flow is too mysterious to comprehend. Little may be extracted from experiments as guides for theoretical treatment. It would be fair to acknowledge that laboratory experiments using various exploitation techniques are almost exhaustive, as far as incompressible flows are concerned. It is lack of a rigorous mathematical framework which hinders progress, because of the difficulties in reconstructing the non-linear characters of fluid dynamics from test data. 
\section{Kinematics of vorticity induction}
The components of Poisson's equation in (\ref{v-poi}) have the explicit expressions
\begin{equation} \label{lapdv}
\begin{split}
\Delta' u & = - \Big( \frac{1}{r}\frac{\partial \zeta}{\partial \theta} - 
\frac{\partial \eta}{\partial z} - \frac{2}{r^2}\:\frac{\partial v}{\partial \theta} \Big), \\
\Delta' v & = - \Big( \frac{\partial \xi}{\partial z} - \frac{\partial \zeta}{\partial r} + \frac{2}{r^2}\:\frac{\partial u}{\partial \theta} \Big), \\
\Delta w & = -\Big( \frac{\eta}{r} + \frac{\partial \eta}{\partial r} - \frac{1}{r}\frac{\partial \xi}{\partial \theta} \Big),
\end{split}
\end{equation}
which are viewed as a system of elliptic equations subject to conditions (\ref{bc}) on the pipe wall, and (\ref{ic}) at the pipe entry. 
Green's function for the Laplacian $\Delta$ can be found by the method of separation of variables. In view of the no-slip condition at the wall, it has the form of
\begin{equation} \label{gg}
\begin{split}
G({\bfx}, {\bfx'}) = \frac{1}{2 \pi }&\Big( \sum_{k=1}^{\infty} \frac{J_0 \big(\lambda_{k} r \big) J_0 \big(\lambda_{k} r' \big)}{\lambda_k \: J^2_1 \big(\lambda_{k})} \:E_0(z,z') \\
&+2\sum_{n=1}^{\infty}\sum_{k=1}^{\infty} \frac{J_n \big(\sigma_{n,k} r \big) J_n \big(\sigma_{n,k} r' \big)}{ \sigma_{n,k} \: (J_{n+1} \big(\sigma_{n,k})\big)^2 } \cos\big(n(\theta{-}\theta')\big) \: E(z,z')\Big),
\end{split}
\end{equation}
where 
\begin{equation*}
\begin{split}
E_0(z,z')=&\:\exp\big({-} \lambda_{k}|z{-}z'|\big) - \exp\big({-} \lambda_{k}|z{+}z'|\big),\\
E(z,z')=&\:\exp\big({-} \sigma_{n,k}|z{-}z'|\big) - \exp\big({-} \sigma_{n,k}|z{+}z'|\big),
\end{split}
\end{equation*}
and the constant, $\lambda_{k}, k=1,2,\cdots,$ is the $k$th positive zero of the Bessel function $J_0(\lambda)=0$, $\sigma_{n,k}$ the $k$th positive zero of $J_n(\sigma)=0$. The zeros, $\sigma_{n,k}$, are larger than $\lambda_1=2.404826$ (see appendix A for some examples). Similarly, Green's function for operator $\Delta'$ is found to be
\begin{equation} \label{gh}
D({\bfx}, {\bfx'}) = \frac{1}{\pi} \sum_{n=1}^{\infty}\sum_{k=1}^{\infty} 
\frac{J_n \big(\sigma_{n,k} r \big) J_n \big(\sigma_{n,k} r' \big)}{ \sigma_{n,k}\:(J_{n+1} \big(\sigma_{n,k})\big)^2} \cos\big(n(\theta{-}\theta')\big)\:E(z,z').
\end{equation}

There exists a rich collection of literature on Bessel functions, see, for example, Watson (1944) and Olver {\it et al.} (2010). 
Bessel functions $J_n(x)$ are entire functions of the argument $x$,
\begin{equation} \label{bess}
J_n(x) = \Big(\frac{x}{2}\Big)^n \sum_{k=0}^{\infty} \frac{(-1)^k\: (x^2/4)^k}{k! \:\Gamma(n+k+1)},\;\;\;n\geq0,
\end{equation}
and
\begin{equation} \label{bess2}
J^2_n(x) = \Big(\frac{x}{2}\Big)^{2n} \sum_{k=0}^{\infty} \frac{(-1)^k \:(x^2/4)^k \:\Gamma(2n+2k+1)}{k! \:(\Gamma(n+k+1))^2\: \Gamma(2n+k+1)}.
\end{equation}
The following recurrence relations are well-known (for $x>0$)
\begin{equation} \label{bess-rec}
J_{n+1}(x)=(2n)J_n(x)/x-J_{n-1}(x)\;\;\; \mbox{and} \;\;\;
 J'_{n}(x)=J_{n-1}(x)-n J_n(x)/x.
\end{equation}
The denominators in the Green functions are derived from the orthogonal relation
\begin{equation*}
2 \int_0^1 x J_n \big(\sigma_{n,k} x \big) J_n \big(\sigma_{n,j} x \big) \rd r = \delta_{kj} \big(\:J_{n+1} \big(\sigma_{n,k} \big) \:\big)^2,
\end{equation*}
where $\delta_{kj}$ is Kronecker's symbol. For the Bessel functions of all order $n\geq 0$,  
\begin{equation*}
\int_0^{\infty} J_{n}(x) \rd x =1,\;\;\;\;\;\;
\mbox{and}\;\;\;\;\;\;
 \big| J_n(x) \big|  \leq 1 \;\;\; \mbox{for} \;\;\; 0 \leq x < \infty.
\end{equation*}
For small arguments, we have
\begin{equation*}
J_0(0)=1,\;\;\;\;\;\; J_{n}(x) \rightarrow x^n\;\;\;\mbox{as}\;\;\; x \rightarrow 0\;\;\; n \geq1.
\end{equation*}
The Bessel functions oscillate for large arguments 
\begin{equation*}
J_n(x) \sim \sqrt{\frac{2}{\pi x}}\cos\Big(\: x - \Big(\frac{n}{2} + \frac{1}{4}\Big) \pi \:\Big)\;\;\;\mbox{as}\;\;\; x \rightarrow \infty.
\end{equation*}
By the series expansion for the Bessel function, we find
\begin{equation} \label{jr}
\frac{J_n(\sigma_{n,k} x) }{x} \leq \frac{\sigma_{n,k}}{2 } \big|J_{n-1}(\sigma_{n,k} x)\big| \leq \frac{\sigma_{n,k}}{2},\;\;\;n\geq1,\;\;\;k\geq1.
\end{equation}
We have carried out some numerical experiments for the double sum 
\begin{equation*}
\sum_{n=1}^{\infty} \sum_{k=1}^{\infty} \frac{J_n \big(\sigma_{n,k} r \big) J_n \big(\sigma_{n,k} r' \big)}{r' \:(J_{n+1} \big(\sigma_{n,k})\big)^2}.
\end{equation*}
For $n,k \sim O(100)$ and $0{<}r,r'{<}1$, all the computations can be done efficiently. There have been no numerical difficulties.

The inversion of the last equation in (\ref{lapdv}) is given by
\begin{equation} \label{w-inv}
\begin{split}
w({\bfx}) & = \int_{\Upomega} r'G(\bfx,\bfx')\Big( \frac{\eta}{r'} + \frac{\partial \eta}{\partial r'} - \frac{1}{r'}\frac{\partial \xi}{\partial \theta'} \Big)(\bfx') \rd \bfx'+ \int_{\Upomega^{\bigotimes}} r'\big[\partial_{z'}G \big](\bfx,\bfy) \;w_e(\bfy) \rd \bfy\\
& = \int_{\Upomega} \Big( {G}\;\eta({\bfx'}) - r'\:(\partial_{r'}G)\;\eta({\bfx'}) + ({\partial_{\theta'}G}) \;\xi({\bfx'}) \Big) \rd \bfx' + W_e(\bfx),
\end{split}
\end{equation}
where the functions,
\begin{equation} \label{ggb}
\partial_{r'}G({\bfx}, {\bfx'})  = \partial G/\partial r',\;\;\;\;\;\;
\partial_{\theta'}G({\bfx}, {\bfx'}) = \partial G/\partial \theta',
\end{equation}
are obtained from integration by parts. The boundary terms vanish in view of the periodic condition for vorticity (\ref{vbc}), and the properties of the Bessel functions. Moreover, the notation $\big[\partial_{z'}G \big]  = \partial G/\partial z'\big|_{z'=0}$, and we have introduced the notation, 
\begin{equation*}
\bfV_e(\bfx) = (U_e\;\;V_e\;\;W_e)(\bfx),
\end{equation*}
to account for the contribution from the velocity at the pipe entry. To continue the process of inversion, we reduce the first two equations in (\ref{lapdv}) to
\begin{equation} \label{lapuv}
\begin{split}
u({\bfx}) & = \int_{\Upomega} \Big( r'(\partial_{z'}D) \: \eta({\bfx'}) - ({\partial_{\theta'}D}) \: \zeta({\bfx'}) \Big)\rd {\bfx}' - 2 \int_{\Upomega} \frac{\partial_{\theta'}D}{{r'}}\:v({\bfx'}) \rd {\bfx}'+U_e(\bfx), \\
& \\
v({\bfx}) & = \int_{\Upomega} r'\Big( (\partial_{z'}D) \: \zeta({\bfx'}) - (\partial_{r'}D) \: \xi({\bfx'})  \Big)\rd {\bfx}' + 2\int_{\Upomega} \frac{\partial_{\theta'}D}{{r'}}\:u({\bfx'}) \rd {\bfx}'+V_e(\bfx),
\end{split}
\end{equation}
where the derivatives on Green's function $D$ are derived in the same manner as those in (\ref{ggb}), and use is made of the velocity periodic condition (\ref{bc2}), and the decay of Green's function at $z' \rightarrow \infty$. In particular, the function $\partial_{\theta'}D/{r'}$ remains regular as $r' {\rightarrow} 0$ by virtue of (\ref{bess}) and (\ref{jr}). The induced velocities by the entry, $u_e$ and $v_e$, are found in an analogous manner to that by $w_e$. In (\ref{w-inv}), we replace the kernel $G$ and $w_e$ by $D$ and $u_e$ so as to obtain $U_e$, 
\begin{equation} \label{gue}
U_e (\bfx)= \int_{\Upomega^{\bigotimes}} r'\big[\partial_{z'}D \big](\bfx,\bfy) \;u_e(\bfy) \rd \bfy,
\end{equation}
where $\big[\partial_{z'}D \big]  = \partial D/\partial z'\big|_{z'=0}$.
The component $V_e$ is obtained analogously. In view of the incompressibility, the induced velocity ${\bfV}_e$ is completely determined and fixed for given geometry. It is evident that the influence of the entry data of finite energy is restricted to a portion of the pipe downstream of the entry, and diminishes at large distance $z \rightarrow \infty$, as both $E_0(z,z')\rightarrow 0$ and $E(z,z')\rightarrow 0$. 

Because of the rotational symmetry from $0$ to $2\pi$, we readily verify that
\begin{equation*} 
\int_{\Upomega}\int_{\Upomega} \frac{\partial_{\theta'}D}{{r'}} \rd \bfx \rd \bfx' =\int_0^1 \!\int_0^1 \int_0^{\infty}\int_0^{\infty} \frac{1}{{r'}}\Big( \int_0^{2 \pi}\!\!\!\int_0^{2 \pi} \partial_{\theta'}D \rd \theta \rd \theta' \Big) \rd z \rd z' \rd r \rd r' = 0.
\end{equation*}
From the properties of the zeros for $J_n(\sigma_{n,k})$, neither $1$ nor $2$ is an eigenvalue of the homogeneous operator $\Delta'u=0$. The equations in (\ref{lapuv}) are a pair of Fredholm integral equations of the second kind with continuous bounded kernels. They may be re-written in vector form for the unknown vector $U=(u\;\;v)$
\begin{equation} \label{lapie}
U + {\mathscr H}*U = {\mathscr K}*\omega+V={\mathscr F},
\end{equation}
where $V{=}(U_e\;\;V_e)$, and the {\it ad hoc} operator $*$ stands for integrals of matrix kernel multiplying a vector function. (It should not be confused with the usual convolution operator.) The domain of integration is clear by considering the kernel and the function; it is over the space $\bfx$. Explicitly, the kernel $\mathscr H=\mathscr H(\bfx,\bfx')$ is a non-zero square matrix with zero trace, 
\begin{equation} \label{lap-h}
{\mathscr H} = \left( \begin{array}{cc}
0 & 2 \:\partial_{\theta'}D/{r'}\\
-2 \:\partial_{\theta'}D/{r'} & 0\\
\end{array}
\right),
\end{equation}
and $\mathscr K=\mathscr K(\bfx,\bfx')$ is $2{\times}3$, 
\begin{equation} \label{lap-k}
{\mathscr K} = \left( \begin{array}{ccc}
0 & r'\partial_{z'}D & -\partial_{\theta'}D\\
-r'\partial_{r'}D & 0 & r'\partial_{z'}D \\
\end{array}
\right).
\end{equation}
Pre-multiplying (\ref{lapie}) by $\mathscr H$, we obtain 
\begin{equation*} 
U + {\mathscr H}*U + {\mathscr H}*{\mathscr H}*U = {\mathscr H}*{\mathscr F}+{\mathscr F}.
\end{equation*}
Carrying out the multiplication once more, we get
\begin{equation*} 
U + {\mathscr H}*{\mathscr H}*{\mathscr H}*U = {\mathscr H}*{\mathscr H}*{\mathscr F}+{\mathscr H}*{\mathscr F}+{\mathscr F}.
\end{equation*}
After $k$-times multiplications, we write the result as
\begin{equation} \label{lapiek}
U  = \sum_{i=1}^k \; \underbrace{ {\mathscr H}*\;\cdots\;{\mathscr H}*}_{i \; \mbox{{\footnotesize fold}} }{\mathscr F} \; + \;  {\mathscr F} \; - \;\underbrace{{\mathscr H}*\cdots*{\mathscr H}*}_{k+1 \; \mbox{{\footnotesize fold}}}U.
\end{equation}
As shown earlier, the boundedness for $\partial_{\theta'}D/{r'}$ implies $\|{\mathscr H}\|_{L^{\infty}} {<} A_0$. Since we are interested in fluid motions of finite energy, we consider $\|\bfu_e\|_{L^{\infty}} {<} B_0$, then the tail on the right in (\ref{lapiek}) vanishes as
\begin{equation*}
B_0A_0^{k+1} \big( \underbrace{I*\cdots*I*}_{k+1 \; \mbox{{\footnotesize fold}}}I \;\big) \leq \frac{B_0A_0^{k+1}} {(k+1)!} \rightarrow 0 \;\;\; \mbox{as} \;\;\; k \rightarrow \infty,
\end{equation*}
where $I$ is the identity matrix. By the Sobolev embedding theorem, we may make reference to the $L^2$-theory for Fredholm integral equations of the second kind (see, for example, Chapter 2 of Tricomi 1957). The operation sum must converge in the limit of $k \rightarrow \infty$.
Let us denote the infinite sum of the multiple operations by $\widetilde{\mathscr H}$, and it is called the resolvent kernel of ${\mathscr H}$. For bounded vorticity, the solution of the integral equations (\ref{lapie}) or (\ref{lapuv}) is given by 
\begin{equation*}
U = {\mathscr K} *\omega + (\widetilde{\mathscr H}* {\mathscr K}) *\omega + V + \widetilde{\mathscr H}* V.
\end{equation*}
Combining the result with (\ref{w-inv}), we write the relation between the velocity and the vorticity as $\bfu={\mathscr V}*\omega+{\mathscr W}*\bfV_e $, or 
\begin{equation} \label{vel-vort}
\begin{split}
\left( \begin{array}{c}
u\\
v\\
w
\end{array}
\right)(\bfx) =   
\int_{\Upomega} &\left( \begin{array}{ccc}
V_{11} & V_{12} & V_{13} \\
V_{21} & V_{22} & V_{23} \\
V_{31} & V_{32} & 0
\end{array} 
\right) (\bfx, \bfx')
\left( \begin{array}{c}
\xi\\
\eta\\
\zeta
\end{array}
\right)(\bfx')\rd \bfx'\\
& \\
& + \int_{\Upomega^{\bigotimes}} \left( \begin{array}{ccc}
W_{11} & W_{12} & 0 \\
W_{21} & W_{22} & 0 \\
0 & 0 & W_{33}
\end{array} 
\right) (\bfx, \bfy)
\left( \begin{array}{c}
\!\!u_e\!\!\\
\!\!v_e\!\!\\
\!\!w_e\!\!
\end{array}
\right)(\bfy)\rd \bfy.
\end{split}
\end{equation} 
This relation holds for every instant of time, but it does not contain time information, since equation (\ref{lapdv}) is elliptic. This is a consequence of the incompressibility hypothesis (\ref{incmp}).
We draw our attention to the fact that the vortices, $\xi$ and $\eta$, do contribute to their velocities $u$ and $v$, in contrast to the velocity induction in the Cartesian co-ordinates. The reason is that $u$ Laplacian is linked to the azimuthal gradient in $v$, which is partly driven by vorticity component $\xi$. When a unit-mass fluid particle at $r=r_0$ from the pipe centre is rotating in the $r{-}\theta$ plane instantaneously, the term $v^2/r_0$ in the $u$-momentum equation in (\ref{cylns}) is the centrifugal acceleration on the particle toward the centre.

It may have been a generalisation of the potential theory or the electromagnetism in classical physics, suggestion has been put forward to represent the velocity field by a solenoidal vector stream function, where $\bfu =\nabla \times \Uppsi$. Instead of (\ref{lapdv}), we have
\begin{equation} \label{phi-omega}
\Delta \Uppsi(\bfx) = - \nabla \times \bfu = - \omega.
\end{equation}
Although this set of lower-order equations appears to be simpler, its general use can be tricky and hence is {\it not} recommended, except for flows in unbounded space $\real^3$. The reason is that no quantitative knowledge of the stream function on solid walls is available; the inversion of (\ref{phi-omega}) cannot be readily obtained. The integral of $\bfu$ contains arbitrary constants, depending on local flow condition. On the other hand, imposing the no-slip condition involves two Green's functions for every component of $\Uppsi$ and hence the evaluation of these Green's functions becomes substantially more complicated.
\section{{Vorticity due to entry flow}}
Because of the specific characters of our co-ordinates, we prefer to write Green's function for the diffusion kernels in (\ref{vdyn}) as 
\begin{equation} \label{vgreen}
\bfH(\bfx,\bfx',t)=Z_1(z,z',t)\:r'\:\bfK(r,r',\theta, \theta',t),
\end{equation}
where $\bfH$ is diagonal, and we denote its components in a row matrix 
\begin{equation*}
\bfH = (\;H\;\;\;H\;\;\;H_3\;).
\end{equation*}
The function,
\begin{equation} \label{z1}
Z_1(z,z',t)=\frac{1}{\sqrt{4 \nu t \pi}} \: \Big\{ \exp\Big(- \frac{(z-z')^2}{4 \nu t}\;\Big) + \exp\Big(- \frac{(z+z')^2}{4 \nu t}\;\Big) 
\Big\},
\end{equation}
is the gradient of the fundamental solution of heat equation in one space dimension ($z\geq 0$). The second part of the Green function satisfying the Neumann homogeneous condition on the wall is written in the diagonal matrix 
\begin{equation*}
\bfK = (\;K\;\;\;K\;\;\;K_3\;).
\end{equation*}
The first two elements are $K=1/\pi+K_0$, where
\begin{equation} \label{green-k12}
K_0=\frac{1}{\pi}  \sum_{n=1}^{\infty}\sum_{m=1}^{\infty} \frac{ \alpha_{n,m}^2 \: J_n \big(\alpha_{n,m} r \big) J_n \big(\alpha_{n,m} r' \big)}{(\alpha_{n,m}^2-n^2) \: (J_n \big(\alpha_{n,m})\big)^2} \cos\big(n(\theta{-}\theta')\big) \: \exp\big({-} \alpha^2_{n,m} \: \nu t\big),
\end{equation}
and
\begin{equation} \label{green-k3}
K_3=\frac{1}{\pi} + \frac{1}{\pi} \;\sum_{m=1}^{\infty}  \frac{J_0 \big(\beta_{m} r \big) J_0 \big(\beta_{k} r' \big)}{J^2_0 \big(\beta_{m})} \:\exp\big({-} \beta^2_{m} \: \nu t\big) + 2K_0.
\end{equation}
The constant, $\beta_{m}, m=1,2,\cdots,$ is the $m$th positive zero of Bessel function $J'_0(\beta)=0$, and $\alpha_{n,m}$ the $m$th positive zero of $J'_n(\alpha)=0$. The zeros, $\alpha_{n,m}$, are larger than $\beta_1=1.841184$ (see appendix A). 

As a first approximation, we investigate the vorticity field caused by the entry vorticity (\ref{vic0}) and, for the time being, we ignore the contributions from the wall. Then the dynamics equations (\ref{vdyn}) can be expressed in integral form:
\begin{equation} \label{vort-j0}
\begin{split}
\omega^{(0)}(\bfx,t)  = 
\int_0^t \!\!\int_{\Upomega^{\bigotimes}} \!\! \bfH_e(\bfx,\bfy,t{-}t') & \upsilon_e
(\bfy,t') \rd \bfy \rd t' \\
& + \int_0^t \!\! \int_{\Upomega}  \bfH (\bfx,\bfx',t{-}t')
\bar{\bf X}^{(0)}(\bfx',t') \rd \bfx' \rd t', 
\end{split}
\end{equation} 
where $\bfH_e$ is a diagonal matrix. We denote the elements by 
\begin{equation*}
\bfH_e = (\;A\;\;\;A\;\;\;A_3\;),
\end{equation*}
and
\begin{equation*}
\bfH_e = - \nu\:r'\: \bfK \;\big[\:Z_1 \:\big]_{z'=0} = - \nu\:r'\: \bfK(r,r',\theta,\theta',t-t') \; Z_0(z,t-t'),
\end{equation*}
where $t>t'$
\begin{equation*}
Z_0(z,t) = \frac{1}{\sqrt{4\nu t \:\pi}} \: \exp\Big(- \frac{z^2}{4 \nu t}\;\Big).
\end{equation*}
As implied in $\bar{\bf X}^{(0)}=\bar{\bf X}^{(0)}(\omega^{(0)})$, the last integral term represents the non-linear interaction in the vorticity field.

By contrast with the vorticity dynamics in the Cartesian co-ordinates, two extra terms arise in (\ref{vdyn}), namely, 
\begin{equation*}
\frac{1}{r^2} \frac{\partial \eta}{\partial \theta}\;\;\;\mbox{and}\;\;\;\frac{1}{r^2} \frac{\partial \xi}{\partial \theta}.
\end{equation*}
We expect that they are invariant with respect to the rotational symmetry. For instance, we obtain the following result by integration by parts: 
\begin{equation} \label{k-red}
\begin{split}
\int_0^1 \frac{1}{r'^2} \Big( \int_0^{2 \pi} & r' K \Big(\frac{\partial \eta} {\partial \theta'}\Big) (r',\theta') \rd \theta' \Big) \rd r'   \\ & =  - \int_{\Upomega^{\bigotimes}} \partial_{\theta'}K (r,r',\theta,\theta') \frac{\eta}{r'} (r',\theta') \rd r' \rd \theta' 
\end{split}
\end{equation}
for fixed time $t'$. We draw our attention to the fact that this reduction is performed independently of the heat kernel $Z_1$. The first two vorticity components in $\omega^{(0)}$ can be reduced to integral equations
\begin{equation} \label{vort-pair-j0}
\begin{split}
\xi^{(0)}(\bfx,t)  =  2 \nu \int_0^t \int_{\Upomega} Z_1\frac{\partial_{\theta'}K}{r'} (\bfx,\bfx',t{-}t') \:\eta^{(0)} (\bfx',t')\rd \bfx' \rd t'
& + A{\starsp}f_e  + H{\starsp}X^{(0)},\\
& \\
\eta^{(0)}(\bfx,t)  = -2 \nu \int_0^t\int_{\Upomega} Z_1\frac{\partial_{\theta'}K}{r'} (\bfx,\bfx',t{-}t') \: \xi^{(0)} (\bfx',t')\rd \bfx' \rd t'
& + A{\starsp}g_e + H{\starsp}Y^{(0)}.
\end{split}
\end{equation}
This is a system of two Volterra-Fredholm integral equations of the second kind (cf. (\ref{lapie})) for the unknowns $(\xi^{(0)},\eta^{(0)})$ over space-time. We note that $K/r'$ is regular at $r'=0$ because $J_1 \propto r$ near $r=0$. The operator ${\star}$ indicates the integration over space-time $(\bfx,t)$. We express the kernel of the system as
\begin{equation*}
 k_0=\left( \begin{array}{cc}
0 & 2 \: \nu \: Z_1\:\partial_{\theta'}K/r' \\
-2 \: \nu \: Z_1\: \partial_{\theta'}K/r' & 0 \\
\end{array} 
\right). 
\end{equation*} 
Evidently, the kernel is a continuous bounded function of $\bfx$ and $t>0$. Let $\widetilde{K}$ be the resolvent kernel of $k_0$. On the basis of standard theory for parabolic differential equations (see, for example, Chapters 1 \& 2 of Friedman 1964), we find that
\begin{equation*}
\big| \widetilde{K} \big| < \frac{C}{(\nu t - \nu t'\:)^{1/2}} \; \exp\Big(- \kappa \frac{(z \pm z')^2}{4 \nu (t -t')} \Big)
\end{equation*}
for a constant $C$, and $0< \kappa < 1$. Denote 
\begin{equation} \label{mtx-h1h2}
\widetilde{R}_1=\widetilde{K} \starsp A,\;\;\;\;\;\;\widetilde{R}_2=\widetilde{K} \starsp H.
\end{equation}
Then we verify that
\begin{equation*}
\big| \widetilde{R}_1 \big| < \frac{C_1}{(\nu t - \nu t'\:)^{1/2}} \; \exp\Big(- \kappa_1 \frac{(z \pm z')^2}{4 \nu (t -t')} \Big),
\end{equation*}
and 
\begin{equation*}
\big| \widetilde{R}_2 \big| < {C_2}(\nu t - \nu t'\:)^{1/2} \; \exp\Big(- \kappa_2 \frac{(z \pm z')^2}{4 \nu (t -t')} \Big),
\end{equation*}
where $C_1$ and $C_2$ are constant, and $0< \kappa_1, \kappa_2 < 1$.
It is well-known that the zeros of the Bessel functions, $\alpha_{n,m}$, are large and positive. Thus, the term in Green's functions $K$ has exponential damping effects in time.
The pair of integral equations can now be solved. The result is combined with the $\zeta$-component to give
\begin{equation} \label{vort-iter-ie}
\begin{split}
\xi^{(0)}  =  &\; A{\starsp}f_e  + \widetilde{R}_1{\starsp}f_e + \widetilde{R}_1{\starsp}g_e + H{\starsp}X^{(0)} + \widetilde{R}_2{\starsp}X^{(0)} + \widetilde{R}_2{\starsp}Y^{(0)},\\
\eta^{(0)}  = &\; A{\starsp}g_e + \widetilde{R}_1{\starsp}g_e + \widetilde{R}_1{\starsp}f_e + H{\starsp}Y^{(0)} + \widetilde{R}_2{\starsp}Y^{(0)} +  \widetilde{R}_2{\starsp}X^{(0)},\\
\zeta^{(0)}  = &\; A_3{\starsp}h_e + H_3{\starsp}Z^{(0)},
\end{split}
\end{equation}
where the entry vorticity in the radial and the azimuthal directions is filtered by the Volterra-Fredholm operator. For example, the second term in the first equation on the right is, for $t>t'>t''$,
\begin{equation*}
\begin{split}
\widetilde{R}_1{\starsp}f_e(\bfx,t) & = \int_{\Upomega_T} \int_{\Upomega^{\bigotimes}_T} \widetilde{K}(\bfx,\bfx',t{-}t')A(\bfx',\bfy,t'{-}t'') f_e(\bfy,t'') \rd \bfy \rd \bfx' \rd t' \rd t''\\
& = \int_{\Upomega^{\bigotimes}_T} \widetilde{R}_1(\bfx,\bfx',t{-}t') f_e(\bfy,t') \rd \bfy \rd t',
\end{split}
\end{equation*}
where we have interchanged the orders of integrals, assuming Fubini's theorem. Similarly, the fifth term is
\begin{equation*}
\begin{split}
\widetilde{R}_2{\starsp}X^{(0)}(\bfx,t) & = \int_{\Upomega_T} \int_{\Upomega_T} \widetilde{K}(\bfx,\bfx',t{-}t')H(\bfx',\bfx'',t'{-}t'') X^{(0)}(\bfx'',t'') \rd \bfx'' \rd t'' \rd \bfx' \rd t' \\
& = \int_{\Upomega_T} \widetilde{R}_2(\bfx,\bfx',t{-}t') X^{(0)}(\bfx',t') \rd \bfx' \rd t'.
\end{split}
\end{equation*}
We write the system (\ref{vort-iter-ie}) in the short-hand form
\begin{equation} \label{vort-g}
\omega^{(0)}= \bfG_e {\starsp} \upsilon_e + \bfG_s {\starsp} {\bf X}^{(0)},
\end{equation}
where the matrix kernels are 
\begin{equation} \label{mtx-ge}
\bfG_e = \left( \begin{array}{ccc}
A+\widetilde{R}_1 & \widetilde{R}_1 & 0 \\
\widetilde{R}_1 & A+\widetilde{R}_1 & 0 \\
0 & 0 & A_3
\end{array} 
\right),
\end{equation}
and
\begin{equation} \label{mtx-gs}
\bfG_s = \left( \begin{array}{ccc}
H+\widetilde{R}_2 & \widetilde{R}_2 & 0 \\
\widetilde{R}_2 & H+\widetilde{R}_2 & 0 \\
0 & 0 & H_3
\end{array} 
\right).
\end{equation} 
From result (\ref{k-red}), definition (\ref{mtx-h1h2}) and properties of kernels $A$ and $H$, we see that the elements of $\bfG_e$ and $\bfG_s$ are proportional to
\begin{equation} \label{time-g0g1}
(\nu t)^{-3/2} \: \exp\Big(- \frac{z^2}{4 \nu t}\Big),\;\;\;\;\;\;\mbox{and}\;\;\;\;\;\;(\nu t)^{-1/2} \: \exp\Big(- \frac{z^2}{4 \nu t}\Big)
\end{equation}
respectively. To save space, we write
\begin{equation*}
\big( \bfu^{(0)},\;\;\omega^{(0)} \big) \;\;\;\mbox{as}\;\;\; \big( \bfu,\;\;\omega \big)^{(0)}.
\end{equation*}
Let us consider $\xi^{(0)}$ component in (\ref{vort-iter-ie}). First
\begin{equation*}
\begin{split}
H{\starsp}X^{(0)}(\bfx,t) & =\int_{\Upomega_T} H(\bfx,\bfx',t{-}t') \Big( (\omega . \nabla \big) u - (\bfu.\nabla) \xi  \Big)^{(0)} (\bfx',t') \rd \bfx' \rd t' \\
& = \int_{\Upomega_T} \nabla_{\bfx'}H \big(\xi v + \xi w - \eta u - \zeta u \big)^{(0)} \rd \bfx' \rd t',
\end{split}
\end{equation*}
where we have performed integrating by parts in the space variable, and made use of the no-slip boundary condition for $\bfu$, the periodic condition for $\bfu$ and $\omega$, zero boundary value for $\bfK$ at $r'{=}0$, and the solenoidal conditions $\nabla.\bfu{=}\nabla.\omega{=}0$. Note that the two terms $\xi u$ from the vorticity convection and stretching cancel each other. By the same token, we find similar expressions for the other two components:
\begin{equation*}
\begin{split}
H{\starsp}Y^{(0)}(\bfx,t) & = \int_{\Upomega_T} \nabla_{\bfx'}H \big(\eta w + \eta u - \zeta v - \xi v \big)^{(0)}  \rd \bfx' \rd t',\\
H_3{\starsp}Z^{(0)}(\bfx,t) & = \int_{\Upomega_T} \nabla_{\bfx'}H_3 \big(\zeta u + \zeta v - \xi w - \eta w \big)^{(0)}  \rd \bfx' \rd t'.
\end{split}
\end{equation*}
Second, consider 
\begin{equation*}
\begin{split}
\widetilde{R}_2{\starsp}(X^{(0)}+Y^{(0)})(\bfx,t)=\int_{\Upomega_T} & \widetilde{R}_2(\bfx,\bfx',t{-}t') \Big( ({\omega} . \nabla \big) u - ({\bf u}.\nabla) \xi  \\
& + ({\omega} . \nabla \big) v - ({\bf u}.\nabla) \eta + (\eta u - \xi v )/r' \Big)^{(0)}  (\bfx',t') \rd \bfx' \rd t'.
\end{split}
\end{equation*}
Carrying out similar expansion, the right-hand side is simplified as
\begin{equation*}
\int_{\Upomega_T} \nabla_{\bfx'}\widetilde{R}_2\Big( (\xi + \eta) w - \xi u - \zeta v \Big)^{(0)}  \rd \bfx' \rd t'
+ \int_{\Upomega_T} \frac{\widetilde{R}_2}{r'} \Big(\eta u - \xi v \Big)^{(0)}  \rd \bfx' \rd t'.
\end{equation*}
There have been some cancellations among the terms arising from the vorticity convection and stretching. We notice that the same reduction can be made for $\eta^{(0)}$ component. Lastly, $\zeta^{(0)}$ component is simplified as
\begin{equation*}
\int_{\Upomega_T} \nabla_{\bfx'}H_3\Big(  \zeta u + \zeta v - (\xi + \eta) w \Big)^{(0)}  \rd \bfx' \rd t'.
\end{equation*}

Thus the system (\ref{vort-g}) can be separated into the individual contributions from the vorticity, and the velocity: 
\begin{equation} \label{vort-f}
\omega^{(0)}= \bfG_e {\starsp} \upsilon_e + \bfC_1 {\starsp} \big(\bfF_1^{(0)}\;\bfu^{(0)}\big) + \bfC_2 {\starsp} \big(\bfF_2^{(0)}\;\bfu^{(0)}\big) + \bfC_3 {\starsp} \big(\bfF_3^{(0)}\; \bfu^{(0)}\big),
\end{equation}
where
\begin{equation} \label{mtx-f1}
\bfF_1^{(0)} = \left(\begin{array}{ccc}
-(\eta+\zeta) & \xi & \xi \\
\eta & -(\zeta+\xi) & \eta \\
\zeta & \zeta & -(\xi+\eta)
\end{array} 
\right)^{(0)},
\end{equation} 
\begin{equation} \label{mtx-f2f3}
\bfF_2^{(0)} = \left(\begin{array}{ccc}
-\zeta & -\zeta & (\xi+\eta) \\
-\zeta & -\zeta & (\xi+\eta)\\
0 & 0 & 0
\end{array} 
\right)^{(0)},\;\;\;\;\;\;
\bfF_3^{(0)} = \left(\begin{array}{ccc}
\eta & -\xi & 0 \\
\eta & -\xi & 0\\
0 & 0 & 0
\end{array} 
\right)^{(0)}.
\end{equation} 
In equation (\ref{vort-f}), the brackets are used to emphasise our conventions that the vorticity and velocity matrix multiplications are performed first. The kernels $\bfC$'s are diagonal, and we summarise their elements in 
\begin{equation} \label{mtx-c}
\begin{split}
\bfC_1 & = \big(\; \nabla_{\bfx'}H\;\;\;\;\;\nabla_{\bfx'}H\;\;\;\;\;\nabla_{\bfx'}H_3 \;\big), \\
\bfC_2 & = \big(\; \nabla_{\bfx'}\widetilde{R}_2 \;\;\;\;\;\nabla_{\bfx'}\widetilde{R}_2 \;\;\;\;\;0 \;\big),\\
\bfC_3 & = \big(\; \widetilde{R}_2/r\;\;\;\;\;\widetilde{R}_2/r\;\;\;\;\;0 \;\big).
\end{split}
\end{equation}
In view of bounds (\ref{time-g0g1}) and the properties of the heat kernel $Z_1$ in Green's function, the time-wise dependence of these elements becomes clear. In particular, we have
\begin{equation} \label{time-cs}
\nabla_{\bfx'}H \; \propto \;\;(\nu t - \nu t' \:)^{-3/2} \: \exp\Big(- \frac{(z \pm z')^2}{4 \nu (t-t')}\Big).
\end{equation}

By virtue of the inverse relation (\ref{vel-vort}), we substitute vorticity $\omega^{(0)}$ for $\bfu^{(0)}$. After straightforward algebraic manipulations, we find that system (\ref{vort-f}) can be expressed as
\begin{equation} \label{vort-q}
\omega^{(0)}= \bfG_e {\starsp} \upsilon_e + \int_{\Upomega_T} \Big( \bfR_0(\bfu_e) +\bfR_1(\omega^{(0)}) \Big)(\bfx,\bfx',t{-}t')\:\omega^{(0)}(\bfx',t') \rd \bfx' \rd t',
\end{equation}
where the elements of $\bfR_0$ and $\bfR_1$ can be found from those $\bfC$'s, $V_{ij}$ and $W_{ij}$. By the estimate (\ref{time-g0g1}), the term associated with the entry vorticity is found to behave like diffusion in one space dimension,
\begin{equation*} 
{\bfG}_e {\starsp} \upsilon_e(\bfx,t)\: \propto (\nu t)^{-1/2} \: \exp\Big(- \frac{z^2}{4 \nu t}\Big).
\end{equation*}
The flow evolution really occurs in three space dimensions because $r$ and $\theta$ vorticity components, $\xi$ and $\eta$, are closely coupled. In terms of the resolvent kernel of $\bfR_0$, this integral equation is further reduced to
\begin{equation} \label{vort-qr}
\omega^{(0)}= \widetilde{\bfG}_e {\starsp} \upsilon_e +\int_{\Upomega_T} \bfQ(\bfx,\bfx',t{-}t';\omega^{(0)}) \omega^{(0)}(\bfx',t') \rd \bfx' \rd t'.
\end{equation}
The last integral is quadratic in vorticity components, $\omega_i^{(0)}$. The kernels $\widetilde{\bfG}_e$ and $\bfQ$ are non-linear functions of the entry velocity $\bfu_e$.

Following our treatment of vorticity development in $\real^3$ (see Chapters 7 and 8 of Lam 2013), we notice that every integral component in (\ref{vort-qr}) may be expressed in the form of
\begin{equation*}
\int_{\Upomega_T} \int_{\Upomega} \sum_{l=1}^3 A_{il}(\bfx,\bfx',\bfy,t{-}t') \omega^{(0)}_l(\bfy,t') \omega^{(0)}(\bfx',t') \rd \bfy \rd \bfx' \rd t',\;\;\;i=1,2,3.
\end{equation*}
In view of (\ref{time-cs}), we have
\begin{equation*} 
\widetilde{\bfG}_e {\starsp} \upsilon_e(\bfx,t)\:\big(\equiv\varpi(\bfx,t)\big)\: \propto (\nu t)^{-3/2} \: \exp\Big(- \frac{z^2}{4 \nu t}\Big).
\end{equation*}
Similarly, coefficient matrix $A_{ij}$ has this identical temporal dependence. Thus the non-linear integral equation (\ref{vort-qr}) has the component form 
\begin{equation} \label{vort0}
\begin{split}
\omega_i^{(0)}(\bfx,t) & =  \varpi_i(\bfx,t)  \\
& + \int_{\Upomega_T} \! \int_{\Upomega} \sum_{l=1}^3 A_{il}(\bfx,\bfx',\bfx'',t{-}t') 
\omega^{(0)}_l(\bfx'',t') \omega^{(0)}(\bfx',t') \rd \bfx'' \rd \bfx' \rd t'.
\end{split}
\end{equation}
These expressions imply an algorithm for similarity reductions; over a small time interval in $\Delta t \in [0,t]$, we multiply the left-hand by $A_{il}$ and $\omega^{(0)}$ for every $i=1,2,3$. The sum of the results is equivalent to the difference of $\omega^{(0)}$ and the initial data $\varpi^{(0)}$. The reductions can be performed as many times as we wish.
Written in vector form, the non-linearity in the integral equation is effectively transformed into an infinite sum: 
\begin{equation} \label{vort0-inf}
\omega^{(0)}(\bfx,t) = \varpi(\bfx,t) + \int_{\Upomega_T} \bfB^{(0)}(\bfx,\bfx',t{-}t') \omega^{(0)}(\bfx',t') \rd \bfx' \rd t' + q(\bfx,t),
\end{equation}
where $q$ is an infinite sequence
\begin{equation*}
q(\bfx,t)=\sum_{m=2}^{\infty} h^{(m)} (\bfx,t),
\end{equation*}
and $h^{(m)}$ stands for $m$-fold integral of $A_{il}$ and $\varpi$. Let $\widetilde{\bfB}^{(0)}$ be the resolvent kernel for $\bfB^{(0)}$. The unknown vorticity in equation (\ref{vort0-inf}) can be recast and expressed as
\begin{equation} \label{vort0-res}
\omega^{(0)}(\bfx,t) = \gamma(\bfx,t) + q(\bfx,t) + \int_{\Upomega_T} \widetilde {\bfB}^{(0)}(\bfx,\bfx',t{-}t')( \gamma+q)(\bfx',t') \rd \bfx' \rd t',
\end{equation}
where $\gamma$ is the Volterra-Fredholm filtered entry data, and is given by
\begin{equation} \label{gamma}
\gamma(\bfx,t) = \varpi(\bfx,t) + \int_{\Upomega_T} \widetilde {\bfB}^{(0)}(\bfx,\bfx',t{-}t')\varpi(\bfx',t') \rd \bfx' \rd t'.
\end{equation}
By the method of successive approximation, equation (\ref{vort0-res}) can be completely solved for any given initial data at pipe entry $\upsilon_e$. We express the solution in terms of the vorticity integral powers $V[\gamma]^k=V[\gamma(\bfx,t)]^k$ in the following convergent series:
\begin{equation} \label{vort-sol-wall0}
 \begin{split}
	\omega^{(0)}(\bfx,t) = & \;\gamma(\bfx,t) \: + 2 \: V[\:\gamma\:]^2 + 10 \: V[\:\gamma\:]^3 + 62 \: V[\:\gamma\:]^4 + 430 \: V[\:\gamma\:]^5 \\ 
	\quad & \\
	& + 3194 \: V[\:\gamma\:]^6 + 24850 \: V[\:\gamma\:]^7 + \; \cdots \cdots \: .
 \end{split}
\end{equation}
It is convenient to write the solution as
\begin{equation} \label{vort0-op}
\omega^{(0)}(\bfx,t) = {\mathscr L} \big(\; \gamma \;\big).
\end{equation}
\section{Flow evolution over space-time}
After a small time $t=t_s$, the flow field defined by (\ref{vort-j0}), or given by local solution (\ref{vort0-op}), occupies the entire interior of the pipe. For convenience, we may re-define the start at time $t=t_s$. Applying Duhamel's principle, the dynamic equations (\ref{vdyn}) can be expressed in integral form ($t>t'$):
\begin{equation} \label{vort-ie}
\begin{split}
\omega(\bfx,t)  = &
\int_0^t  \!\!\int_{\Upomega^{\bigotimes}}  \!\! \bfH_e(\bfx,\bfy,t{-}t') \upsilon_e
(\bfy,t') \rd \bfy \rd t' + \int_{\Upomega}  \bfH (\bfx,\bfx',t) \omega_s(\bfx') \rd \bfx'  \\
& + \int_0^t \!\! \int_{\partial\Upomega} \!\! \bfH_b (\bfx,\bfz,t{-}t')
\varphi_b(\bfz,t')\rd \bfz \rd t'\\
& + \int_0^t \!\! \int_{\Upomega}  \bfH (\bfx,\bfx',t{-}t')
\bar{\bf X}(\bfx',t') \rd \bfx' \rd t'\;\Big(=I_e + I_s + I_b + \Phi\Big),\;t>t'.
\end{split}
\end{equation} 
This equation comprises all the vorticity contributions of the initial-boundary value problem (cf. (\ref{vort-j0})). 
\subsection{Initial vorticity}
As in many physical problems where parabolic differential equations are applicable, the second term on the right is the contribution from the initial vorticity. If the interior fluid is set into motion impulsively from rest, instead of the void vorticity (\ref{vic0}), we must specify the initial distribution over the whole pipe wall in the form of a vortex sheet. The strength of the sheet is difficult to quantify without approximation. By considering the local-in-time shears on the wall due to the entry vorticity (\ref{vort-sol-wall0}), we prescribe the initial data for vorticity as
\begin{equation} \label{vick}
\omega_s(r,\theta,z>0)=(\;\xi_s\;\;\;\eta_s\;\;\;\zeta_s\;)(r,\theta,z>0)=\omega^{(0)}\;(r,\theta,z>0,t=t_s).
\end{equation}
An alternative, which is popular in numerical computations and circumvents the uncertainty in the initial data, is to disturb an {\it existing} Hagen-Poiseuille parabolic profile over a part of the pipe length. Written in dimensionless form, parabolic distribution, $w(r){=}1{-}r^2$, is widely used. The mathematical model is periodically extended over the whole length of the pipe. The two underlying assumptions in this model are that, Hagen-Poiseuille flow is absolutely steady in the complete pipe, and the parabolic profile exists at arbitrary Reynolds number. However, as an initial-boundary value problem, a fluid motion may start from an arbitrary initial profile. The dependence of the initial condition cannot be ignored, as the initial data dominates the subsequent flow development. In fact, both the assumptions can hardly be justified (see \S \ref{flowregimes}).
\subsection{Wall vorticity}
The third integral states the contribution of the vorticity production from the boundary; it is calculated by means of time-dependent integral relation
\begin{equation} \label{vort-wall}
\int_{\partial\Upomega_T}  \bfH_b (\bfx,\bfz,t{-}t')
\varphi_b(\bfz,t')\rd \bfz \rd t', 
\end{equation} 
where the vorticity derivative in the $r$-direction at wall is evaluated from
\begin{equation} \label{dvwall}
\varphi_b(\bfz,t) = \frac{\partial \omega(\bfx,t)} {\partial r} \Big|_{r=1},
\end{equation}
and the kernel is given by
\begin{equation} \label{wall-kernel}
\bfH_b = \nu \: Z_1 \; [\:r' \bfK \:]_{r'=1}  = \nu \: Z_1(z,z',t) \: \bfK'(r,\theta,z,\theta',z',t).
\end{equation}
The diagonal kernel $\bfH_b$ has elements
\begin{equation*}
\bfH_b=(\;B\;\;\;B\;\;\;B_3\;).
\end{equation*}
Similarly, function $\bfK' = (\:K'\;\;\;K'\;\;\;K'_3\:)$, where $K'=1/\pi+K'_0$, and 
\begin{equation*} 
K'_0=\frac{1}{\pi}  \sum_{n=1}^{\infty}\sum_{m=1}^{\infty} \frac{ \alpha_{n,m}^2 \: J_n \big(\alpha_{n,m} r \big)}{(\alpha_{n,m}^2-n^2) \: J_n \big(\alpha_{n,m}\big)} \cos\big(n(\theta{-}\theta')\big) \: \exp\big({-} \alpha^2_{n,m} \: \nu t\big),
\end{equation*}
and
\begin{equation*} 
K'_3=\frac{1}{\pi} + \frac{1}{\pi} \;\sum_{m=1}^{\infty}  \frac{J_0 \big(\beta_{m} r \big)}{J_0 \big(\beta_{m}\big)} \:\exp\big({-} \beta^2_{m} \: \nu t\big) + 2K'_0.
\end{equation*}
The analytic form of the kernel Green function indicates that the wall layer is determined by every part of the flow though the influence at the far end of the pipe is clearly insignificant, specifically for small data. For initial flow of arbitrary size, our method of solution takes the global characters into account since the vorticity field so solved ($\varphi_b$) contains all the interaction among the shears of different scales over the complete pipe length. In particular, the wall term retains the essential structure of the vorticity solution (\ref{vort-sol-wall0}). The dependence of the kernel $\bfH_b$ on viscosity 
$\nu$ shows that it is absolutely essential to consider the viscous effect. On the other hand, the wall vorticity gradient illustrates the fact that the pipe flow field is dynamic in nature. Modelling pipe flow in terms of steady Hagen-Poiseuille flow of parabolic profile ought to be a crude approximation for the study of the dynamics; it would be misleading to consider the axial symmetrical flow over a significant part of the pipe length, particularly over Reynolds number range of turbulence transition.

We call the flow structure at the solid surfaces as a wall layer in contrast to a boundary layer as the latter suggests that Prandtl's boundary layer approximations are involved. Evidently, the flow is everywhere fully viscous, and there does not exist a clear-cut ``free-stream" where the flow may be considered as uniform and inviscid. Equation (\ref{ppoi}) implies that the pressure must be strongly dependent on the instantaneous values of the velocity gradients. The shears themselves evolve in the transport process of convection and diffusion. 

One particular feature of the wall vorticity is rather interesting. Suppose that the motion is set up impulsively, and we then maintain a sustained supply of the constant flow rate through the entry. The local vorticity diffuses out of the wall into the interior in the same way as heat is released from a surface. The velocity gradient normal to the wall contains a natural decay, i.e., the factor $\exp(-\alpha^2_{n,m}\: \nu t)$. For fluids of small viscosity, the decay near the entry appears to be time-insensitive, and the diffusion may proceed linearly even in the presence of the wall. This is of course a misinterpretation, because the main vorticity dynamics is determined by $\varphi_b$ which is non-linear and unsteady. In particular, if vorticity $\omega(\bfx,t)$ has a complex structure like in a turbulence, the flow in the wall layer can be readily understood, even though experimental measurements may be difficult to conduct.
\subsection*{Solution by time-marching}
We solve the vorticity evolution (\ref{vort-ie}) by considering the following iterations: For 
$j=1,2,3,\cdots$,
\begin{equation} \label{vort-iter}
\omega^{(j)}(\bfx,t)  = I_e(\bfx,t) + I^{(j{-}1)}_s(\bfx,t) + I_b(\bfx,t) + \Phi (\omega^{(j)})(\bfx,t),
\end{equation}
where the initial data are given by
\begin{equation} \label{vort0j}
I_s^{(j-1)}(\bfx,t) = \int_{\Upomega} \bfH (\bfx,\bfx',t)
\omega^{(j{-}1)}(\bfx')\rd \bfx'.
\end{equation} 
The wall vorticity gradient (\ref{dvwall}) is written as
\begin{equation} \label{vort0-wall}
I_b(\bfx,t) = \int_0^t \int_{\partial\Upomega} \bfH_b (\bfx,\bfz,t{-}t')
\varphi_b(\bfz,t')\rd \bfz \rd t',
\end{equation} 
where the starting value may be taken from the result of $\omega^{(0)}$, 
\begin{equation*}
\varphi_b^{(0)}=\frac{\partial \omega^{(0)}}{\partial r}\Big|_{r=1}(\theta,z,t)=(\:p_b\;\;\;q_b\;\;\;r_b\:)(\theta,z,t).
\end{equation*}
\subsection*{Solution $\omega^{(1)}$}
The new vorticity evolves according to
\begin{equation} \label{vort-iter1}
\begin{split}
\omega^{(1)}(\bfx,t) = I_e(\bfx,t) + I_s^{(0)}(\bfx,t) &+ \int_{\partial\Upomega_T} \bfH_b (\bfx,\bfz,t{-}t')\varphi_b(\bfz,t')\rd \bfz \rd t' \\
& \hspace{5mm}+\int_{\Upomega_T} \!\!\bfH (\bfx,\bfx',t{-}t')
\bar{\bf X}^{(1)}(\bfx',t') \rd \bfx' \rd t'.
\end{split}
\end{equation}
The updated pair of integral equations (\ref{vort-pair-j0}) now read
\begin{equation} \label{vort-pair-j1}
\begin{split}
\xi^{(1)}(\bfx,t)  =  2 \nu \int_0^t \int_{\Upomega} & Z_1\frac{\partial_{\theta'}K}{r'} (\bfx,\bfx',t{-}t') \:\eta^{(1)} (\bfx',t')\rd \bfx' \rd t' \\
& + A{\starsp}f_e + H{\starsp}\xi^{(0)}_s + B{\starsp}p_b + H{\starsp}X^{(1)},\\
\eta^{(1)}(\bfx,t)  = -2 \nu \int_0^t\int_{\Upomega} & Z_1\frac{\partial_{\theta'}K}{r'} (\bfx,\bfx',t{-}t') \: \xi^{(1)} (\bfx',t')\rd \bfx' \rd t' \\
& + A{\starsp}g_e + H{\starsp}\eta^{(0)}_s + B{\starsp}q_b + H{\starsp}Y^{(1)}.
\end{split}
\end{equation}

Evidently, the reduction of the integral involving $\bar{\bf X}^{(1)}$ can be carried out in the same manner as we did for $\bar{\bf X}^{(0)}$ (cf. (\ref{vort-j0})). There is no point to repeat the detailed procedures.
Equation (\ref{vort-f}) takes the revised form of
\begin{equation} \label{vort-iter-f1}
\begin{split}
\omega^{(1)} = \bfG_e {\starsp} \upsilon_e & + \bfG_s {\starsp} \omega^{(0)}_s + \bfG_b{\starsp} \varphi_b^{(0)}  \\
& + \bfC_1 {\starsp} \big(\bfF_1^{(1)}\:\bfu^{(1)}\big) + \bfC_2 {\starsp} \big(\bfF_2^{(1)}\:\bfu^{(1)}\big) + \bfC_3 {\starsp} \big(\bfF_3^{(1)}\:\bfu^{(1)}\big),
\end{split}
\end{equation}
where $\bfG_s$ is given by (\ref{mtx-gs}), and
\begin{equation} \label{mtx-gb}
\bfG_b = \left( \begin{array}{ccc}
B+\widetilde{K}_b & \widetilde{K}_b & 0 \\
\widetilde{K}_b & B+\widetilde{K}_b & 0 \\
0 & 0 & B_3
\end{array} 
\right),
\end{equation}
and $\widetilde{K}_b = \widetilde{K} {\starsp} B$. The matrices, $\bfF_1$, $\bfF_2$ and $\bfF_3$, are still given by (\ref{mtx-f1}) to (\ref{mtx-f2f3}) with the current value of vorticity $\omega^{(1)}$. Since
\begin{equation*}
B,\;\;B_3  \; \propto \; (\nu t)^{-1/2} \: \exp\Big(- \frac{z^2}{4 \nu t}\Big),
\end{equation*}
we calculate that
\begin{equation*}
\widetilde{K}_b \; \propto \; (\nu t- \nu t' \: )^{1/2} \: \exp\Big(- \kappa \frac{(z \pm z')^2}{4 \nu (t-t')}\Big)
\end{equation*}
for $0< \kappa <1$. Equation (\ref{vort-q}) is revised as
\begin{equation} \label{vort-q1w}
\begin{split}
\omega^{(1)}(\bfx,t) = & {\bfG}_e {\starsp} \upsilon_e + {\bfG}_s {\starsp} \omega^{(0)}_s + \bfG_b{\starsp} \varphi_b^{(0)} \\
& + \int_{\Upomega_T} \Big( {\bfR}_0(\bfu_e) + {\bfR}_1(\omega^{(1)}) \Big)(\bfx,\bfx',t{-}t')\: \omega^{(1)}(\bfx',t') \rd \bfx' \rd t'.
\end{split}
\end{equation}
Although we have not yet had the full structure of the wall layer, we may infer from this result that it is dominated by the global vorticity field. Essentially, the last two integrals contain the quadratic non-linearity. The analogous equation to (\ref{vort-qr}) has the expression
\begin{equation} \label{vort-qr1}
\begin{split}
\omega^{(1)}= \widetilde{\bfG}_e {\starsp} \upsilon_e + \widetilde{\bfG}_s {\starsp} \omega^{(0)}_s &+ \widetilde\bfG_b{\starsp} \varphi_b^{(0)}\\
& +\int_{\Upomega_T} \bfQ(\bfx,\bfx',t{-}t';\omega^{(1)}) \omega^{(1)}(\bfx',t') \rd \bfx' \rd t'.
\end{split}
\end{equation}
In view of the velocity-vorticity relation (\ref{vel-vort}), the non-linear integral equation for the vorticity can be derived:
\begin{equation} \label{vort1}
\begin{split}
\omega_i^{(1)}(\bfx,t)&= \; \varpi^{(0)}_i(\bfx,t) \\
& + \int_{\Upomega_T} \! \int_{\Upomega} \sum_{l=1}^3 A^{(1)}_{il}(\bfx,\bfx',\bfx'',t{-}t') 
\omega^{(1)}_l(\bfx'',t') \omega^{(1)}(\bfx',t') \rd \bfx'' \rd \bfx' \rd t',
\end{split}
\end{equation}
where it is clear that the revised $\varpi$,
\begin{equation*}
\varpi^{(0)}(\bfx,t) =  \varpi(\bfx,t) + \widetilde{\bfG}_s {\starsp} \omega^{(0)}_s(\bfx,t) + \widetilde\bfG_b{\starsp} \varphi_b^{(0)}(\bfx,t),
\end{equation*}
takes into account the initial vorticity and the vorticity production at the wall. Accordingly, the function $\gamma$ can be updated to
\begin{equation} \label{gamma1}
\gamma^{(1)}(\bfx,t) = \varpi^{(0)}(\bfx,t) + \int_{\Upomega_T} \widetilde {\bfB}^{(1)}(\bfx,\bfx',t{-}t')\varpi^{(0)}(\bfx',t') \rd \bfx' \rd t'.
\end{equation}
The solution of (\ref{vort1}) is given by
\begin{equation*}
\omega_0^{(1)}(\bfx,t) = {\mathscr L} \big(\; \gamma^{(1)} \;\big),
\end{equation*}
where the subscript refers to $\varphi_b^{(0)}$.
Next we perturb the wall vorticity gradient $\varphi_b^{(0)}$ by a small amount to 
\begin{equation*}
\varphi_b^{(1)} = \varphi_b^{(0)} + \delta\varphi_b.
\end{equation*}
From the dependence of the wall vorticity given in $\varpi^{(0)}$, we have
\begin{equation*}
\frac{\partial \omega_0^{(1)} }{\partial \varphi_b^{(0)}}(\bfx,t) \approx \widetilde\bfG_b(\bfx,t) +  \int_{\Upomega_T} \widetilde {\bfB}^{(1)}(\bfx,\bfx',t{-}t')\widetilde\bfG_b(\bfx',t') \rd \bfx' \rd t' < \infty.
\end{equation*}
Thus the starting wall vorticity $\varphi_b^{(0)}$ is replaced by 
\begin{equation*}
\varphi_b^{(1)}=\frac{\partial \omega_0^{(1)}}{\partial r}\Big|_{r=1}(\theta,z,t)=(\:p_b\;\;\;q_b\;\;\;r_b\:)(\theta,z,t).
\end{equation*}
With the updated wall vorticity gradient, we solve equation (\ref{vort-iter1}) to obtain the updated solution $\omega_1^{(1)}$. This is nothing more than a Newton-Raphson procedure. We continue our iterations until
\begin{equation*}
\max_{(\bfx,t)}\:\big|\:\omega_{k+1}^{(1)} - \omega_{k}^{(1)} \:\big| < \varepsilon,\;\;\;k=1,2,3, \cdots,
\end{equation*}
where $\varepsilon$ is a prescribed tolerance at $k=M$. The final vorticity solution is denoted by $\omega_{M}^{(1)}(\bfx,t)=\omega^{(1)}(\bfx,t)$.
\subsection*{Solution $\omega^{(j)}$}
Once $\omega^{(1)}$ has been determined, we march into time $t+\delta t$. By shifting the time origin, we carry out the similar calculations for $\omega^{(2)}$. First, we update $I_s^{(0)}$ in (\ref{vort-iter1}) to $I_s^{(1)}$. In view of the Newton-Raphson iteration, the wall vorticity and hence $\omega^{(2)}$ can be found. The solution may be written as
$\omega^{(2)}(\bfx,t) = {\mathscr L} \big(\; \gamma^{(2)} \;\big)$.
To continue the time-marching, suppose that we have the solution for $\omega^{(j-1)}$,
\begin{equation*}
\omega^{(j-1)}(\bfx,t) = {\mathscr L} \big(\; \gamma^{(j-1)} \;\big).
\end{equation*}
The evolutional solution, $\omega^{(j)}$, is determined from the current equation (\ref{vort-iter}):
\begin{equation} \label{vort-iterj}
\begin{split}
\omega^{(j)}(\bfx,t) =  I_e(\bfx,t) + I_s^{(j{-}1)}(\bfx,t) &+ \int_{\partial\Upomega_T} \bfH_b (\bfx,\bfz,t{-}t')
\varphi_b^{(j{-}1)}(\bfz,t')\rd \bfz \rd t' \\
& \hspace{5mm} +\int_{\Upomega_T} \!\!\bfH (\bfx,\bfx',t{-}t')
\bar{\bf X}^{(j)}(\bfx',t') \rd \bfx' \rd t',
\end{split}
\end{equation}
where $I_s^{(j{-}1)}$ is the vorticity initialisation from $\omega^{(j{-}1)}$, and the starting wall value is given by
\begin{equation*}
\varphi_b^{(j{-}1)}=\frac{\partial \omega^{(j-1)}}{\partial r}\Big|_{r=1}.
\end{equation*}
The first two functions on the right in {\ref{vort-iterj}) are known. The third one may be determined by iterations. Now equation (\ref{vort-iter-f1}) becomes
\begin{equation} \label{vort-iter-fj}
\begin{split}
\omega^{(j)} = \bfG_e {\starsp} \upsilon_e & + \bfG_s {\starsp} \omega^{(j{-}1)}_s + \bfG_b{\starsp} \varphi_b^{(j{-}1)}  \\
& + \bfC_1 {\starsp} \big(\bfF_1^{(j)}\:\bfu^{(j)}\big) + \bfC_2 {\starsp} \big(\bfF_2^{(j)}\:\bfu^{(j)}\big) + \bfC_3 {\starsp} \big(\bfF_3^{(j)}\:\bfu^{(j)}\big).
\end{split}
\end{equation}
Since the structure of this non-linear equation is identical to (\ref{vort-iter-f1}) for the case $j{=}1$, in view of the velocity-vorticity link (\ref{vel-vort}), we convert it into the vorticity quadratic form, i.e.,
\begin{equation} \label{vortj}
\begin{split}
\omega_i^{(j)}(\bfx,t)&= \; \varpi^{(j-1)}_i(\bfx,t) \\
& + \int_{\Upomega_T} \! \int_{\Upomega} \sum_{l=1}^3 A^{(j)}_{il}(\bfx,\bfx',\bfx'',t{-}t') 
\omega^{(j)}_l(\bfx'',t') \omega^{(j)}(\bfx',t') \rd \bfx'' \rd \bfx' \rd t'
\end{split}
\end{equation}
(cf. (\ref{vort0})). Thus the function $\gamma^{(j-1)}$ can be updated:  
\begin{equation} \label{gammaj}
\gamma^{(j)}(\bfx,t) = \varpi^{(j-1)}(\bfx,t) + \int_{\Upomega_T} \widetilde {\bfB}^{(j)}(\bfx,\bfx',t,t')\varpi^{(j-1)}(\bfx',t') \rd \bfx' \rd t'.
\end{equation}
We iterate on the wall vorticity gradient, using the Newton-Raphson method to obtain $\omega^{(j)}$. 
By the method of similarity reduction and successive approximations, we assert that the solution of (\ref{vort-iter}) is given by
\begin{equation*}
\omega^{(j)}(\bfx,t) = {\mathscr L} \big(\; \gamma^{(j)} \;\big),\;\;\;j=1,2,3,\cdots.
\end{equation*}
The solution is unique and regular for any $t{>}0$.
\section{Gradated flow regimes} \label{flowregimes}
\subsection*{Mathematical description of turbulence}
We have demonstrated that the vorticity dynamics governed by the non-linear integral equation (\ref{vort-ie}) is globally well-posed. Its solution can be represented in the series
\begin{equation} \label{vort-sol}
 \begin{split}
	\omega(\bfx,t) =  \;\widetilde\gamma(\bfx,t) \:& + 2 \: V[\:\widetilde\gamma\:]^2 + 10 \: V[\:\widetilde\gamma\:]^3 + 62 \: V[\:\widetilde\gamma\:]^4 
	+ 430 \: V[\:\widetilde\gamma\:]^5  \\
	\quad & \\
	\quad & + 3194 \: V[\:\widetilde\gamma\:]^6 + 24850 \: V[\:\widetilde\gamma\:]^7 + 199910 \: V[\:\widetilde\gamma\:]^8  \\
	 \quad & \\
	 \quad & + 1649350 \: V[\:\widetilde\gamma\:]^9 + 13879538 \: V[\:\widetilde\gamma\:]^{10} + \; \cdots \cdots \: ,
 \end{split}
\end{equation}
where $\widetilde\gamma=\widetilde\gamma(\upsilon_e)$. Function $\upsilon_e$ or $\omega_e$ is the vorticity value at pipe entry and may be called {\it the input vorticity} as it completely specifies the flow evolution. 

A vorticity power term, $V[ \;\widetilde{\gamma}\;]^k{=}V[ \;\widetilde{\gamma}\;]^k(\bfx,t)$, generally represents the $k$-fold integral convolution on the initial data $\widetilde{\gamma}$. Over space-time, the convolution effect cultivates a ``vortice"; the higher $k$, the smaller the vorticity scale because of the heat kernel $Z_1$. The coefficient integers form an integer sequence, and the integral convolutions obey certain combinatorics rules.\footnote{This is sequence {\ttfamily A107841} in The On-Line Encyclopaedia of Integer Sequences {\ttfamily www.oeis.org}} For every additional term in the series, the number of vortices is increased roughly by an order of magnitude compared to the preceding scale. The solution (\ref{vort-sol}) designates turbulence in the sense that the vorticity field is composed of interacting shears in numerous scales. The solution can be computed to an arbitrary degree of accuracy, at least in principle. The numerical values obtained at specific space-time locations represent the mean quantities in fluid motion. In reality, physics demands a demarcation on the minimum allowable scales. It is a fact from experiment that turbulent motions observed in nature and laboratory can be described by ensemble-averages and statistically invariant fluctuations (an assumption by Reynolds),
\begin{equation*}
\bfu(\bfx,t) = \bar{\bfu}(\bfx,t)+\bfu'(\bfx,t),\;\;\;\;\;\;\overline{\bfu'}=0,
\end{equation*}
where an overbear denotes the mean or the ensemble-average. Both properties are repeatable in practice, as long as the experimental samples are sufficiently large, so that the theory of large numbers is applicable. The missing random characters in real turbulence is out of the question in the current continuum approach. Because of the viscous diffusion, we may examine the origin of the randomness in terms of the Maxwell-Boltzmann kinetic theory of gases. The law of energy conservation dictates that the loss of the kinetic energy via viscous dissipation equals the increase in the internal energy of fluid's molecules. The diffusive process on the shears of dissipative scale must be irregular because the thermal energy of the molecules fluctuates randomly (Lam 2013). From the application point of view, we notice that the structure of the wall vorticity $\psi_b$ resembles many features in turbulece.

To preserve the generality of solution (\ref{vort-sol}), we shall not, at this stage of analysis, classify flows into particular categories, such as rolls, streaks, hairpins and stationary or travelling waves though these dynamic flow features are basic building blocks for complex flow fields at large Reynolds number. The process of initiation, sustainment and diversification must be history-dependent and scale-specific, at least during the early phase of flow evolution. 

Given entry data (\ref{ic}), the magnitude and distribution of the input vorticity $\omega_e$ determines the subsequent dynamic flow development. In general, a characteristic velocity at every axial location throughout the pipe prior to transition must be the function of entry velocity $\bfu_e$. As the local flow is in evolution, the local velocity is evolving and hence cannot be constant. In the present paper, we define the Reynolds number as
\begin{equation} \label{rey}
Re = \frac{\bar{w}  d} {\nu},
\end{equation}
where $\bar{w}=\bar{w}(\bfu_e)$ is the mean velocity, and $d$ the pipe diameter. This is a textbook definition for Hagen-Poiseuille parabolic profile, 
\begin{equation} \label{hpf1}
{w}/{w_{\mbox{\scriptsize max}}}=1-r^2,
\end{equation}
and $\bar{w}=w_{\mbox{\scriptsize max}}/2$. In practice, it is extremely difficult, if ever possible, to generate and maintain the parabolic velocity from pipe entry to end. Most pipe flow experiments are carried out in the condition of constant stream-wise pressure gradient or of constant mass-flow, for flow starting from specific inlet design to minimise the local disturbances. Velocity measurements in the vicinity of pipe entry indicate that the local profiles are {\it not} self-similar. A Hagen-Poiseuille parabolic profile or the fully developed pipe flow may only be realised at downstream distance of order of $100{-}1000$ diameters for {\it small initial data} or {\it lower to moderate} $Re$. 

We would like to emphasise the fact that, the similarity rule implied in relation (\ref{rey}) is valid strictly for the given distribution $\bfu_e$. Care is needed in the interpretation of conclusions drawn from different experimental measurements, particularly for those on the nature of laminar-turbulent transition. 

\subsection{Uni-directional approximation for entry flow}
The velocity components in the radial and circumferential directions are assumed as identically zero over a small time interval $t \in [0,T_\epsilon]$. The axial component is treated as a function of the variables $r$ and $t$ only, i.e., $w=w(r,t)$, and the pressure varies with the axial variable, $p=p(z,t)$. The differential equation is much simplified as 
\begin{equation} \label{hpfns-ibp}
\frac{\partial w} {\partial t} - \nu \Big( \frac{\partial^2 w}{\partial r^2} + \frac{1}{r} \frac{\partial w}{\partial r} \Big) = \nu P_0(z,t),\;\;\;w(1)=0,\;\;\;w(r,0) = w_e(r),
\end{equation}
where $P_0=-(\partial p/ \partial z)/\mu$ denotes the axial pressure gradient. The solution of the initial-boundary value problem can be expressed in terms of the Fourier-Bessel series (cf. Lam 2015),
\begin{equation} \label{hpf-ibp}
w(r,t)=\int_0^1 w_e(r') G_0(r,r',t) \rd r' + \nu \int_0^t \int_0^1 P_0(z,\tau) G_0(r,r',t{-}\tau) \rd r' \rd \tau,
\end{equation}
where Green's function is given by
\begin{equation} \label{hpf-green}
G_0(r,r',t)=2 \:r'\:\sum_{n=1}^{\infty} \; \exp \big(- \lambda^2_n  \nu t \big) \: \frac{J_0 \big(\lambda_n r \big) J_0 \big(\lambda_n r' \big)}{J^2_1 \big(\lambda_n  \big)}.
\end{equation}
This is the one-dimension version of (\ref{green-k12})-(\ref{green-k3}). Consider the case of constant pressure gradient and the velocity is normalised by $P_0/4$. The scaled velocity is computed from
\begin{equation} \label{uniprof}
\begin{split}
w^*(r,t) & =  \int_0^1 \!w^*_e(r') G_0(r,r',t) \rd r' + (4 \nu t) \! \int_0^1 \!\! \int_0^1 G_0(r,r',t(1{-}\tau)\big) \rd r' \rd \tau \\
& =w_I+w_G.
\end{split}
\end{equation}
(If the initial data $w_e{=}0$ or the motion is assumed to start impulsively from rest, see Szyma{\'n}ski 1932 and \S4.3 of Batchelor 1967, the second term produces a parabolic profile in the limit of $\nu t {\rightarrow} \infty$, i.e., $w_{\mbox{\scriptsize max}}{=}1$ in the distribution (\ref{hpf1}).) 
In figures \ref{we1} and \ref{we2}, we present the time evolution of two constant entry profiles. Briefly, the actual long-time evolution depends on the relative strength of the two contributions; the first term involving initial data $w^*_e$ may well be dominant in the solution (\ref{uniprof}) and hence renders the parabolic profile irrelevant.
 
\begin{figure}[ht] \centering
  {\includegraphics[keepaspectratio, height=14cm,width=14cm]{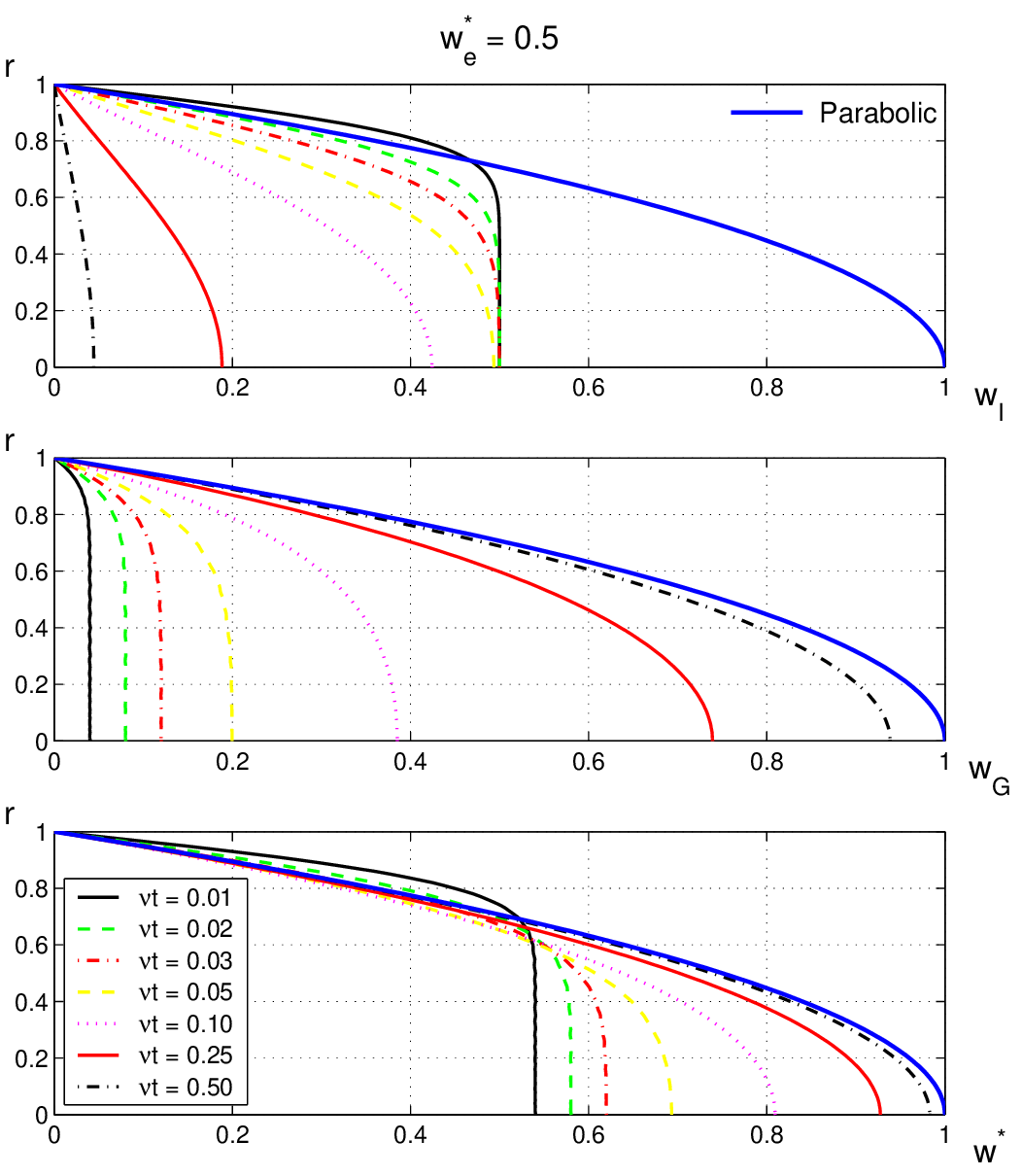}}
  \vspace{1mm}
  \caption{Influence of initial data on flow regime near pipe entry by theory of uni-directional flow. For scaled entry velocity: constant distributions $w^*_e = 0.5$ and $w^*_e = 4.0$. The pipe velocity is given by (\ref{uniprof}) for several fixed values $\nu t$. As time $\nu t$ increases, the initial velocity decays and the wall layer grows in thickness, while the flow driven by the constant pressure gradient $\partial p/\partial z$ builds up. The combined profile $w^*$ is shown in the last row. As $\nu t {\rightarrow} \infty$, the initial influence diminishes, and the pressure gradient becomes dominant, until the total profile approaches to the Hagen-Poiseuille parabolic distribution (\ref{hpf1}). For {\it small initial data}, our solution explains why the fully developed pipe flow at large $\nu t$ has a parabolic velocity distribution. The developed flow may be treated as history-independent after sufficiently long time. }\label{we1} 
\end{figure}
\begin{figure}[ht] \centering
  {\includegraphics[keepaspectratio, height=14cm,width=14cm]{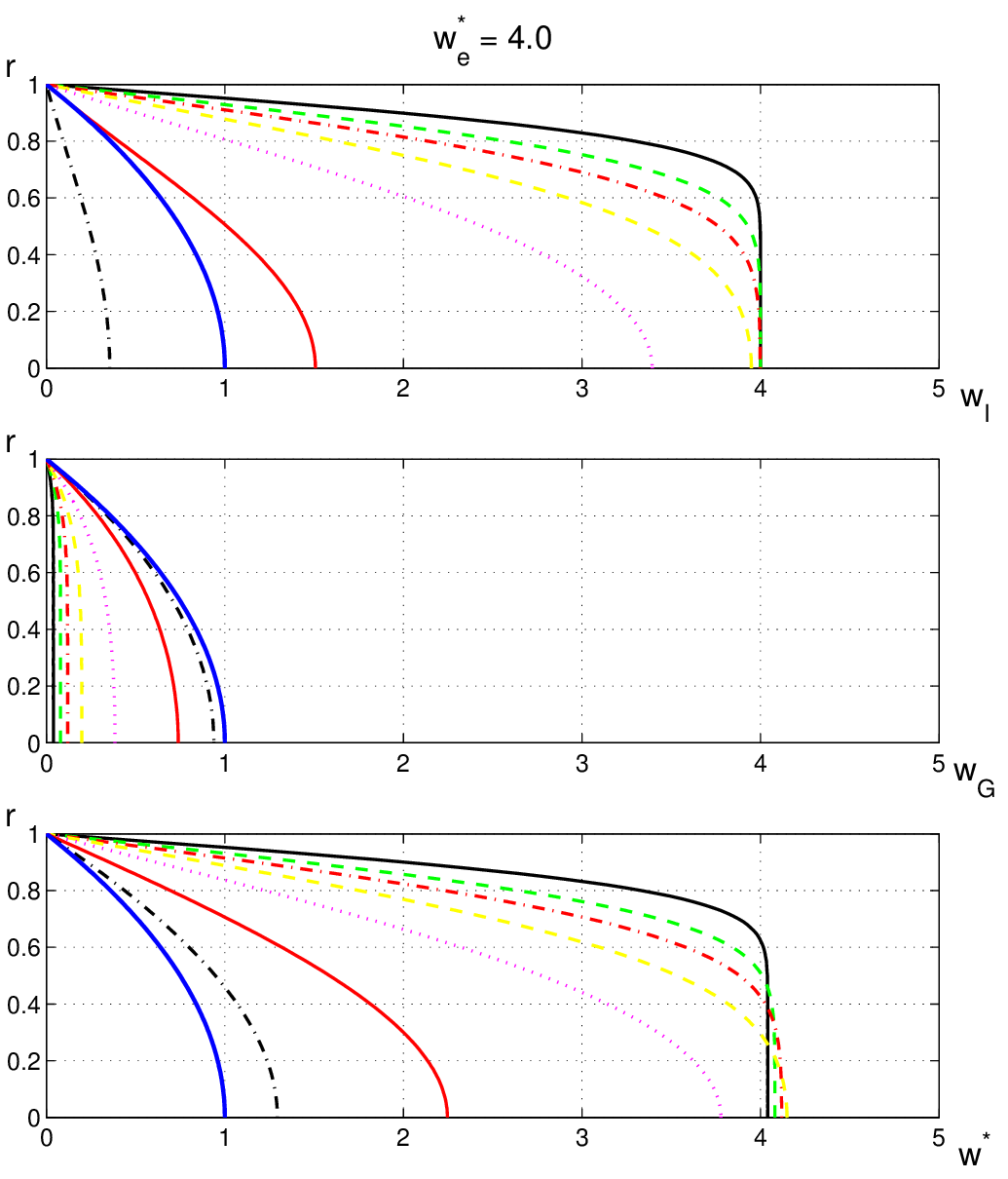}}
  \vspace{1mm}
  \caption{Importance of the initial unsteady phase in pipe flow. The magnitude of entry velocity is increased to $w^*_e = 4.0$, so that the developing pipe profiles are far from parabolic over almost all time. For flow development at high Reynolds number, whether the local flow is laminar or turbulent depends strongly on the initial data. The Hagen-Poiseuille profile (\ref{hpf1}) is just a particular case of {\it developing} flows, that can be generated in long circular pipes.}\label{we2} 
\end{figure}

To relate our theory of uni-directional flow (\ref{hpf-ibp}) in the vicinity of pipe entrance to experiments, we set
\begin{equation*}
\frac{\nu t}{a^2} = \frac{x_{\mbox{\scriptsize expt}}}{a}\frac{1}{R_r} =  \frac{x_{\mbox{\scriptsize eff}}}{d} \frac{1}{Re}, 
\end{equation*}
where the quantity $x_{\mbox{\scriptsize eff}}$ is an effective distance measured from an origin of zero viscous thickness at the theoretical ``entrance", and $R_r$ is the Reynolds number based on pipe radius. It is known from experimental work (Smith 1960; Pfenninger 1961; Tritton 1976; Mullin 2011) that pipe inlets are streamlined designs to minimise irregularity in starting flow. The present approximate theory does not model any specific designs, and hence the velocity profile is assumed to start some distance upstream of the entrance. At entry location, $z{=}0$, the local velocity then satisfies the no-slip condition. This practice results in a shift in quantity $\nu t/a^2$. Comparisons in figures \ref{exptv} and \ref{exptvbl} show that the {\it unsteady solution} (\ref{uniprof}) is very satisfactory over a wide range of $\nu t$, despite of the use of flow conditioners in the experiments. The initial-boundary value formulation of the Navier-Stokes equations must be well-founded. Even in the present approximation, the solution is an excellent description of the initial pipe flow development over substantial entry-length. 

\begin{figure}[ht] \centering
  {\includegraphics[keepaspectratio, height=10cm,width=12cm]{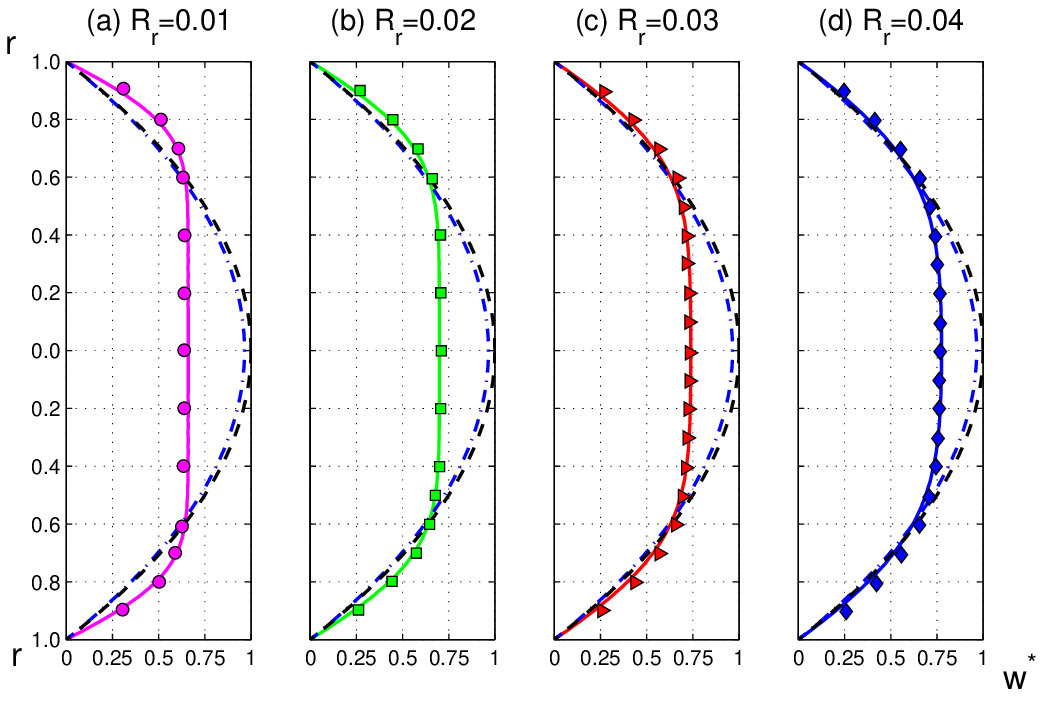}}
  \vspace{1mm}
  \caption{Theory of uni-directional flow for initial evolution in circular pipes. Experimental results are averaged values of the data given in Smith (1960) at $R_r = \bar{w} a/\nu$ $ = 0.01,\:0.02,\:0.03,\:0.04$. Note that the data were taken from three collections of experiments, where the test conditions were very different, and, in particular, the detailed inlet geometries were not fully reported, while the initiation flows were specifically tailored. In the theory, the entry velocity is considered as a constant value $w^*_e = 0.6$. The full lines are calculations at $\nu t/a^2 = 0.015, \:0.025, \:0.035, \:0.045$ for a pipe of radius unity with sharp-edged inlet. The broken blue line is a solution at $\nu t/a^2 =0.25$ which indicates a possible decay of the inlet flow. As $\nu t \rightarrow \infty$, the flow can be treated as fully developed and tends to parabolic profile (\ref{hpf1}) (broken black line). The two broken lines are given for reference only, as the uni-directional flow is supposed to exist in a small time interval from the start of motion.}\label{exptv} 
\end{figure}
\begin{figure}[ht] \centering
  {\includegraphics[keepaspectratio, height=10cm,width=12cm]{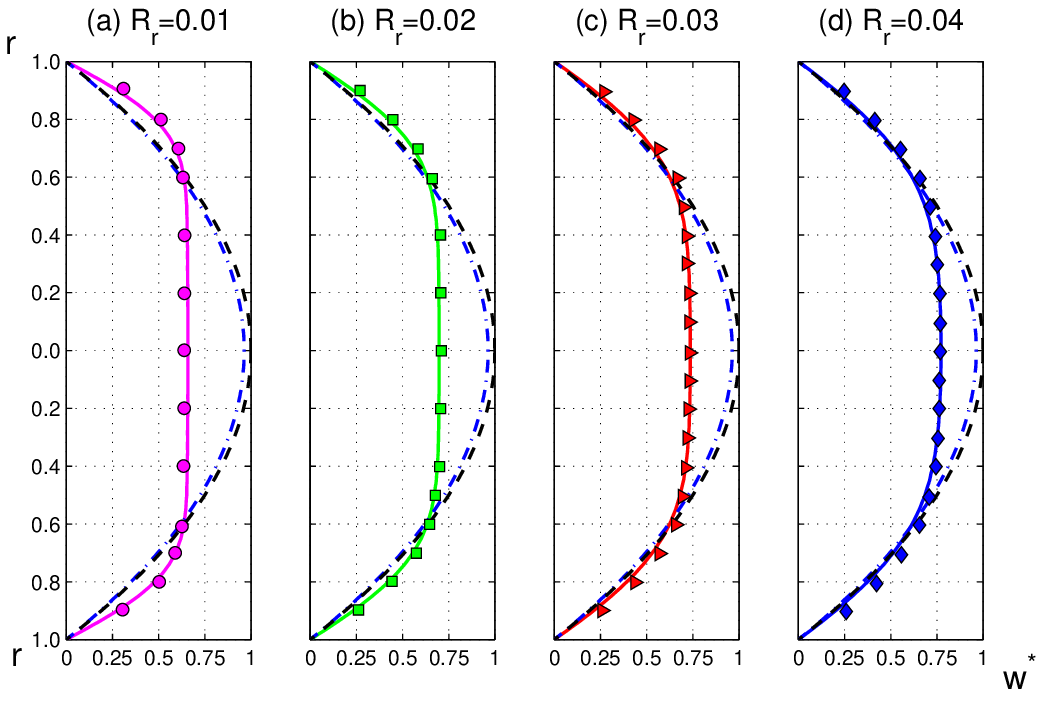}}
  \vspace{1mm}
  \caption{Theory of uni-directional flow for initial evolution in circular pipes. See previous plot for experimental data and theoretical computations. The entry velocity is a boundary layer $w^*_e = 0.6 (1 - r)^{1/10}$. Compared to the case of constant velocity, the differences are negligible, as both theoretical profiles are assumed to start somewhere upstream of the pipe entry. The starting profiles at $z{=}0$ are almost identical. From the experiment point of view, it is much easier to produce a constant velocity in the vicinity of the pipe inlet.}\label{exptvbl} 
\end{figure}

Our chief interest here is to study how turbulence transition initiates naturally. We wish to elucidate the way in which a streamlined flow structure evolves into a complex vorticity field, {\it in the absence} of any disturbances. The experiments by Durst \& Uensal (2006) are a synopsis which demonstrates many important aspects of the natural transition in well-crafted test ambience (figure \ref{vprof}). 

\begin{figure}[t] \centering
  {\includegraphics[keepaspectratio,height=10cm,width=12cm]{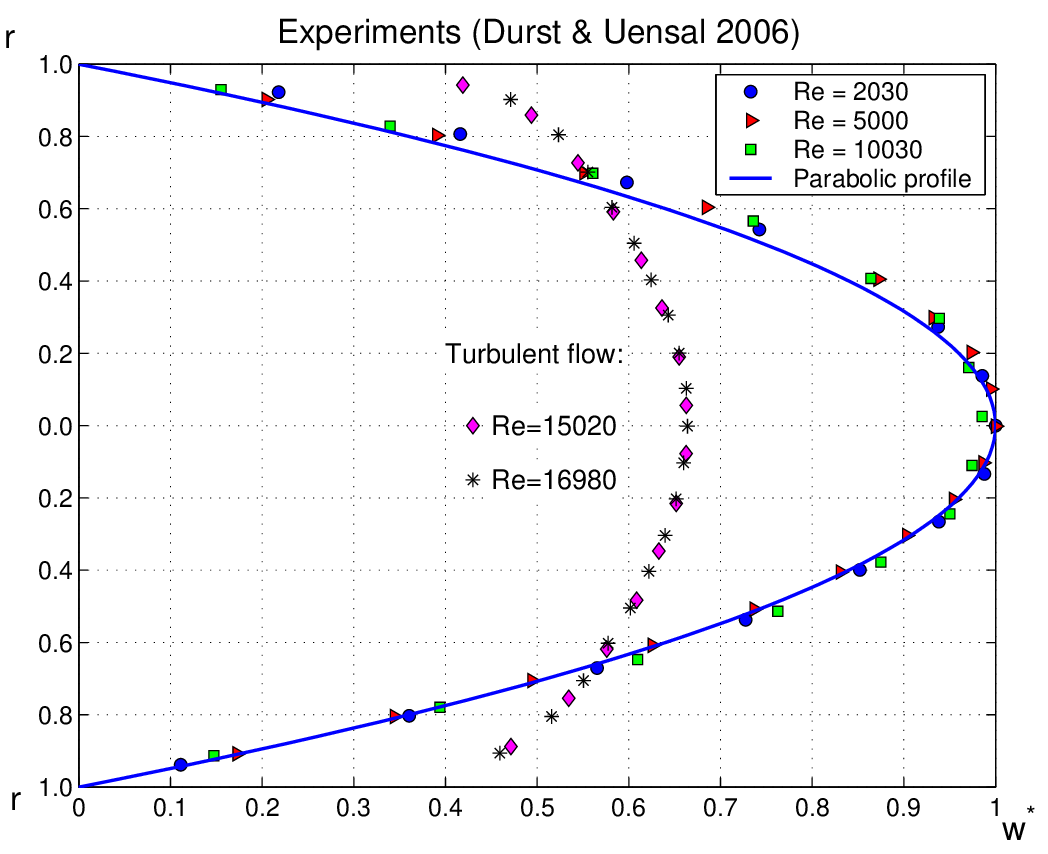}}
  \vspace{1mm}
\caption{In this experimental investigation for natural transition in a pipe of diameter $15\texttt{mm}$ and a length $x=10\texttt{m}$, constant mass flows via a feedback control were maintained. To a good approximation, the Hagen-Poiseuille profile was achieved at every Reynolds number $Re {\leq} 10000$. Transition was detected at $Re_{tr}{\approx} 13000$. At $Re{=}2030,5000,10030,13000$, the values of $\nu t/a^2$ were $1.313, 0.533, 0.266, 0.205$ respectively. All the profiles were measured at the pipe exit. The manner in which the flow was initiated suggested that the profile at the inlet ought to be either a constant or a boundary layer. It is seen that, at $Re{=}10030$, the measured flow was an {\it evolving} profile. As $Re$ was increased to $Re_{tr}$, the transitional profile was certainly non-self-similar. The test data at $Re{=}15020$ and $16980$ showed that the process of losing laminarity initiated and finalised at locations upstream of the exit. In fact, the local flows over transitional process need not to be any particular type, as long as its progenitor at start is sizeable such that the non-linearity in the governing equations is evoked.} \label{vprof} 
\end{figure}
\subsection*{Distinct flow stages}
In his pipe experiments (Reynolds 1883), Reynolds established that there are two fundamental categories of flows: laminar and turbulent. Laminar flows are characterised by well-defined streamlines. Turbulence is observed on macroscopic scale as a fluctuating random motion consisting of various flow scales. The intermediate flow stage between these two flow states during flow evolution is known as transition. The process of transition from the laminar state to turbulence attracts most attention. The reverse scenario is known as re-laminarisation. Intuitively, turbulence transition must merge the streamlined structures and the growing gradations of finer-scale flows. As established from experiments, transition appears to initiate in localised regions in intermittent manners. There are patches of irregularly fluctuating flows during evolution; these are commonly referred as turbulent puffs, or slugs, or spots, which are often recognised in wall-bounded shear flows. 

For a flow of small Reynolds number or a fluid motion with small initial data, the vorticity distribution is given by the leading term $\widetilde{\gamma}$ in (\ref{vort-sol}), and the flow is said to be laminar. The dissipation of the kinetic energy is dominated by diffusion on large-scale. As $Re$ is slightly increased, the vorticity solution is considered to consist of the two-term expansion,
\begin{equation*}
	\omega(\bfx,t) = \widetilde{\gamma} + 2\: V[\:\widetilde{\gamma}\:]^2.
\end{equation*}
The flow can still be regarded as laminar, and evidently the flow field has a large-scale structure because the space-time convolutions on $\widetilde{\gamma}$ produce flow scales comparable to the initial data. However, some increased degree of unsteadiness is detectable in the flow field, due to the interaction among the eddies. If the Reynolds number is further increased, the instantaneous vorticity field is best described by
\begin{equation*}
	\omega(\bfx,t) = \widetilde{\gamma} + 2\: V[\:\widetilde{\gamma}\:]^2 + 10 \: V[\:\widetilde{\gamma}\:]^3,
\end{equation*}
or possibly, by 
\begin{equation*}
	\omega(\bfx,t) = \widetilde{\gamma} + 2\: V[\:\widetilde{\gamma}\:]^2 + 10 \: V[\:\widetilde{\gamma}\:]^3 + 62\: V[\:\widetilde{\gamma}\:]^4.
\end{equation*}
The water in the pipe now consists of multiple eddies of smaller scales, which induce pronounced fluctuations in the flow. In practice, Reynolds did not observe the vorticity field but the velocity field. In view of the induction effect of the Green functions, the velocity field, must show a space-time intermittent character, due to the localised vorticity patches by the interaction of the interwoven $\widetilde{\gamma}$ and the wall layer (cf. $\psi_b$). He referred to the regions of the velocity induced by the vorticity concentrations as flashes (now commonly known as turbulent puffs and slugs). The whole flow field evolves into a structure which looks like a random motion. The Green's functions take into account the downstream transitional or weakly turbulent flow states which in turn randomise the upstream localised vorticity slugs. At a higher $Re$, the number of smaller eddies grows rapidly so that the flow becomes 
\begin{equation*}
\omega(\bfx,t) = \widetilde{\gamma} + 2\: V[\:\widetilde{\gamma}\:]^2 + 10 \: V[\:\widetilde{\gamma}\:]^3 + 62\: V[\:\widetilde{\gamma}\:]^4 + 430\: V[\:\widetilde{\gamma}\:]^5 + 3194 \: V[\:\widetilde\gamma\:]^6.
\end{equation*}
The vorticity field must appear to be chaotic and irregular, due to the strong interactions among the vorticity eddies. As $Re$ is increased further, numerous gradated eddies grow out of the non-linearity, and viscous dissipation intensifies. The randomness mechanism becomes marked and effective. The whole flow soon becomes chaotic, as well as, random as it contains a multitude of eddies in distinct scales. Beyond this stage, the flow field develops into full turbulence. 

{\itshape Turbulence transition in pipe flow is an intrinsic, three-dimensional, non-linear process during the evolution. As the Reynolds number increases, the orderly streamlined structure of initial laminarity is successively modified by the proliferated eddies of progressively smaller size, until the entire structure is dissolved into a composition of fluctuating shears. Depending on given initial data, there exists a Re-number range in which the non-linearity first becomes conspicuous. The process of transition can be an abrupt event or can develop intermittently over a large section of space-time.}
\subsection{Turbulent puffs and slugs}
Local turbulent puffs are highly interactive vorticity due to the coupling of the $r$ and $\theta$ components, as given in the governing equations (\ref{vdyn}). The resolvent kernel rooted in (\ref{vort-iter-ie}) or more appropriately (\ref{vort-pair-j1}) accumulates the components into local vorticity blobs. The convolutions are stronger for the vorticity in the wall layer. The concentrated vortices is advected downstream by the mean stream-wise velocity $w$, so that they become elongated and spread out, forming localised vorticity as puffs. The process of proliferation by the non-linearity tends to weaken the local vorticity structures, and enhances the local vorticity scales. Turbulent slugs are amalgamated puffs due to the scale multiplications. The puffs and slugs exist locally over part of the pipe flow. It is reasonable to expect that the resulting motion incurs intermittent characters, due to the induction of the vorticity in the whole pipe. 

Turbulent puffs refer to localised patches of disordered flow. Their legnths are typically $10{-}20$ pipe-diameter long. They are normally found at $1600{<}Re{<}2500$ in good experimental set-ups and advect downstream at $90{-}95\%$ of the pipe mean velocity (see, for instance, Wygnanski \& Champagne 1973). They are often called ``equilibrium puffs" because they have structured flow features, which do not seem to vary among different experiments. In most experiments, the deductions of their lifetime, or the travelling speed must be inferred from the ensemble-averages of the measurements. Our theory suggests that, in a typical puff, the local vorticity consists of flow scales in the order of 
\begin{equation*}
10\; [\;\widetilde{\gamma}\;]^3\;\;\; {\mbox{to}}\;\;\; 430 \;[\;\widetilde{\gamma}\;]^5\;\;\;\;\;\;\mbox{(turbulent puffs)}.
\end{equation*}
The mean sizes of the vortices ought to be moderate, and are well-above the dissipative scales. Thus, the turbulent puffs are not random in general, as compared to turbulence. But their dynamic behaviours do require statistical descriptions as the number of vorticity scales involved is already rather large. It is essential to notice that these puffs are not strong and abundant enough to trigger transition, as they represent the transitional flow structures between the streamlined laminar flow, and the weak turbulence. As the initial data become stronger or the Reynolds number becomes larger, more vortices of smaller scales will be created out of the non-linearity. As indicated in the re-laminarisation of the puffs by reducing the Reynolds number (Peixinho \& Mullin 2006), the puff structures in the subsequent development remain in an identifiable fashion and they do not become chaotic. In particular, the test results suggest that the decay of the puff follows an exponential rate in space and in time (a typical smoothing effect due to heat kernel). The flow visualisation highlights the existence of $two$ longitudinal vortices preceding the appearance of the puff flow at Reynolds number $Re{=}1750$. It can be interpreted from many experimental data that, at the Reynolds number $Re {<} 1750$, the pipe flow has large-scale non-chaotic structures (\S 2 of Mullin 2011). The two-terms solution of (\ref{vort-sol}) precisely describes the flow composition in the early evolution. 

Turbulent slugs are related to the next flow scales in the vorticity hierarchy of the non-linearity. They can be described as vorticity patches with the scales from the vorticity powers 
\begin{equation*}
430\; [\;\widetilde{\gamma}\;]^5\;\;\; {\mbox{to}}\;\;\; 3194 \;[\;\widetilde{\gamma}\;]^6\;\;\;\;\;\;\mbox{(turbulent slugs)}.
\end{equation*}
The slugs are more space-filling, as demonstrated in flow visualisations. The fluid motions in slugs show strong agitations, and contain localised hysteresis, due partly to the downstream flow effects. However, at Reynolds number $Re{>}3000$, the flow state of initialisation turbulence is characterised by the presence of slugs. The statistical properties of slugs ought to have similarities typically found in turbulence. However, the flow inside slugs is unlikely to resemble full turbulence. Unfortunately, current experimental techniques may not be able to uncover every detail of slug flow, given the presence of numerous vortices. Hence use of statistical means for transition study is inevitable.
\subsection{The intrinsic transition}
The ultimate aim of transition investigation is to identify the process, or the mechanism, which instigates the numerous vortices of various scales, apparently out of streamlined progenitor flow. In reality, experimental evidence shows a great variety in the actual processes, depending on the types of the flow in question (free shear flows, boundary layers, confined flows in pipes or closed channels).
The geometry of circular pipe is simple for experiments, though we have seen that the analytic solutions are much more complicated compared to free shear flows in unbounded space. It is known that pipe flow development is experiment-dependent because of varied test conditions. Transition phenomena in different test environments can be compared only if the disturbance levels are kept below some threshold values (see below \S~\ref{disfree}). Should disturbances be significant and potentially aberrant, flow gradations to turbulence, as an initial-boundary problem of the equations of motion, may take divergent paths and case-dependent routes. We call the laminar-turbulent transition which is free of influence of disturbances {\it the intrinsic transition}. 

The proliferation of vortices does not necessarily occur over the {\it complete} cross-sectional area at any fixed stream-wise $z$ location. The intermittency is a natural phenomenon during flow evolution and transition. Whether local intermittent characters will persist into the subsequent turbulence depends on the initial data and the vicinity flow conditions.

In practice, velocity profiles other than the self-similar distribution can also be generated and maintained. In figure \ref{pfprof} we plot and compare selected test data of Pfenninger (1961).
\begin{figure}[t] \centering
  {\includegraphics[keepaspectratio,height=10cm,width=12cm]{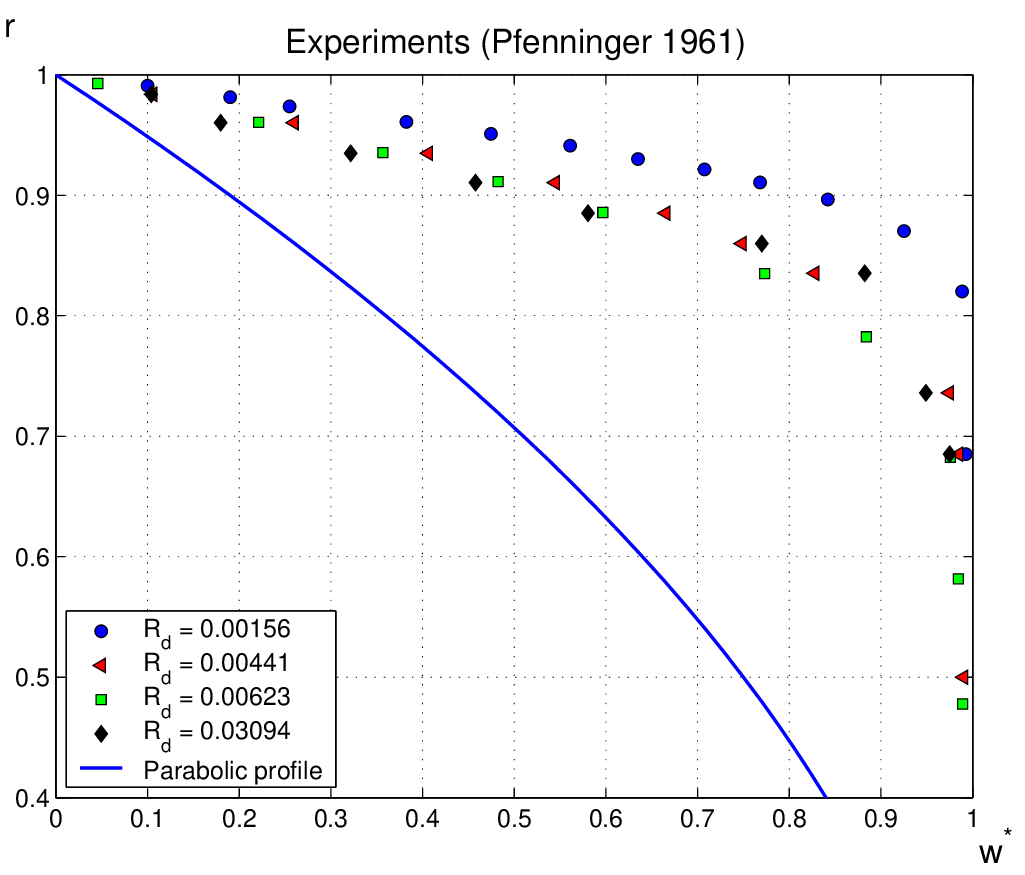}}
  \vspace{1mm}
\caption{Measured velocity profiles in accelerating flows, in favourable pressure gradients, at various $R_d=4({x^*}/d)/{Re}$, where pipe diameter $d=50.8\texttt{mm}$, and $x^*$ is the length of an equivalent straight pipe with zero boundary layer thickness at the entry. For the values of $R_d$, the axial distances from the entry are $1.37\texttt{m}$, $3.045\texttt{m}$, $5.94\texttt{m}$ and $11.88\texttt{m}$ respectively. The velocity profiles substantially deviate from the parabolic profile over the whole pipe length. The main experiments were conducted in accelerating flows, likely involving strong swirls. The energy of the flows in the favourable pressure gradients was much higher than the parabolic profile in constant pressure gradient at the same Reynolds number.} \label{pfprof} 
\end{figure}
It is comprehensible that delayed transition up to $Re{\sim}10^{5}$ was achieved. These tests ought to be assessed as dedicated flow controls to postpone turbulence transition in the same way that suction is applied to laminar boundary layers in high Reynolds number. In theory, the test $\upsilon_e$ must be so powerful that only the first two terms in the vorticity series are adequate to describe the initial evolution over the complete pipe length. On the other hand, there does not exist experimental evidence to indicate that fully developed Hagen-Poiseuille flow of parabolic profile remains laminar at arbitrary Reynolds number.

On the basis of the outcomes from credible constant-mass-flow-rate experiments (see, for example, Darbyshire \& Mullin 1995; Durst \& Uensal 2006), a number of conclusions are compatible with the regularity theory of the Navier-Stokes equations. Arrays of test techniques, including suction, jets, ring blockage, and iris diaphragm, have been utilised to instigate finite perturbations into flows at high Reynolds number. The established results showed that transition is relatively insensitive to the means of forcing which are the representation of initial data. In particular, {\it good repeatability in transition measurements} has been confirmed in both natural and forced scenarios. One may opt the term $\nabla u_i$ in every component of (\ref{ns}) as the unknown in view of $\bfu {=} \Delta^{-1} \omega$, the equations then become a set of {\it quasi-linear} differential equations, instead of non-linear ones. Such a view may be instructive in understanding the quantitative repeatability of the test results, that have been collected at the Reynolds numbers preceding the full-blown random effect due to viscous dissipation. The differential operators, ${\partial_t} {-}  \nu \Delta$ and ${\partial_t} {-}  \nu \Delta'$ in (\ref{vdyn}), are the stereotype of continuum diffusion process for finite $\nu{>}0$. The flow evolution governed by the Navier-Stokes equations does not behave like certain lower-order dynamical systems which are sensitive to initial conditions.

A theoretical development of pipe flow transition is to identify particular numerical solutions known as travelling waves. They are identified as unstable solutions of the governing equations, and are stream-wise vortices propagating with constant phase speed. Specifically, the wavy topology has been considered to originate from saddle-node bifurcations of the stability boundaries in the Reynolds number range $770 {\leq} Re {\leq} 1250$ (Faisst \& Eckhardt 2003; Wedin \& Kerswell 2004). However, {\it the Navier-Stokes equations are deterministic and globally well-posed; exact solutions of the governing equations do not bifurcate in space or in time}. 
\section{Experimental evidence for disturbance-free transition} \label{disfree}
One of the key outcomes from the present theoretical analysis is that the laminar-turbulent transition is a direct consequence of the non-linear term $(\bfu.\nabla)\bfu$ in the equations of motion. All the integral power terms in (\ref{vort-sol}) are emanated from the non-linearity. Thus, our analysis asserts that the intrinsic transition occurs even in the complete absence of disturbances, infinitesimal or finite. Since fluid motions are an initial-boundary value problem of the Navier-Stokes equations, initial and boundary conditions make allowance for any disturbances.
\subsection*{Microscopic fluctuation and turbulence}
One may contemplate that, microscopic fluctuations are the internal disturbances that play a role in turbulence transition. However, molecular agitations are incessant in {\it all} types of fluid motions; the fluctuations exist in laminar, transitional, as well as turbulent flows. In fact, their existence was first discovered in fluids at rest (Brownian motion). Turbulent flows in finite-energy motion are dynamic in nature, and hence do not persist indefinitely. Experiments demonstrate that turbulence in pipe flow undergoes reverse transition during flow decay when a large share of the energy has been dissipated. In most laboratory tests under standard temperature and pressure, any temperature increase from the viscous dissipation of flow energy is not critical enough to alter the mean molecular speed according to kinetic theory of gases (Maxwell 1867; Boltzmann 1906). For incompressible flows, the internal energy of fluid's molecules has a primarily identical Boltzmann distribution in all the three flow stages. If the microscopic fluctuations are responsible for the transition from laminar to turbulent state, in the absence of any macroscopic disturbances, are they the cause for the process of reverse transition?

Nevertheless, the microscopic fluctuations are identified as the origin of the random characters in turbulence in incompressible flows (Lam 2013). Viscous dissipation on the copious eddies of small scales expedites entropy production in order to restore the flows to the equilibrium of shear-free state. The diffusion of individual small-scale eddies is a stochastic process in space and in time, as fluid's thermal energy fluctuates. Fluid motions on the continuum inhere the microscopic randomness through the diffusive dissipation process.
\subsection*{Environmental disturbances}
Environmental disturbances are the external disturbances, such as free-stream turbulence, acoustic noise, mechanical vibrations of test apparatus. They can be specified in the initial data (\ref{ic}) or (\ref{dvic}), and in the boundary value (\ref{bc}). In principle, their effect on transition has been fully taken into account in the equations of motion. Specifically, any background disturbance at the far end of the pipe is fully enumerated in the Green function (\ref{vgreen}) or (\ref{vort-wall}) which encloses the entire pipe interior. The Fourier-Bessel expansion in the function explicates the flow field as a space-bounded wave-like structure with diminishing magnitude as $\nu t {\rightarrow} \infty$. The inherent analytic characters of the solution precisely explain many wavy patterns observed in flow visualisations (see, for instance, van Dyke 1982; and Samimy {\it et al.} 2003). 

In the classic experiments of Schubauer \& Skramstad (1947) for transition flows on a flat-plate in zero pressure gradient, it was demonstrated that the transition location is independent of the free-stream turbulence once the level dropped below a threshold about $0.08\%$. The turbulence level is defined as
\begin{equation*}
T_{{\bfu}} = \Big( \sqrt{\big(\:\overline{{u'}^{2}}+\overline{{v'}^{2}}+\overline{{w'}^{2}}\:\big)/3} \; \Big)/{\bfu}_{\infty},
\end{equation*}
expressed as a percentage. It was speculated that the acoustic excitation in the tunnel might be responsible for the transition. Additional tests showed that the transition was delayed if the environmental noise was carefully regulated (Wells 1967). In particular, when the noise level was below $90$\texttt{db}, the transition was basically insensitive to the ambient disturbances as long as $T_{{\bfu}} {<} 0.1\%$. In figure \ref{xtr} we summarise the transition location as a function of the free-stream turbulence with reduced acoustic excitation. The limiting case of zero disturbances can be obtained by extrapolation. In fact, these limits have been drawn in the early reviews, see, for example,  figure~10 of Tani (1969) or figure~16(a) of Narasimha (1985). However, the significance of these limits has not been realised, due probably to lack of theoretical evidence at the time. Instead, the occurrence of the natural transition was attributed to residue or background disturbances, such as derivations from inlet flow, thermal convection due to temperature differences, or imperfections in test apparatus. Nevertheless, all these perturbation irregularities can be minimised by means of control, and in principle they are best regarded as extraneous for fluid motions.
\begin{figure}[t] \centering
  {\includegraphics[keepaspectratio,height=10cm, width=12cm]{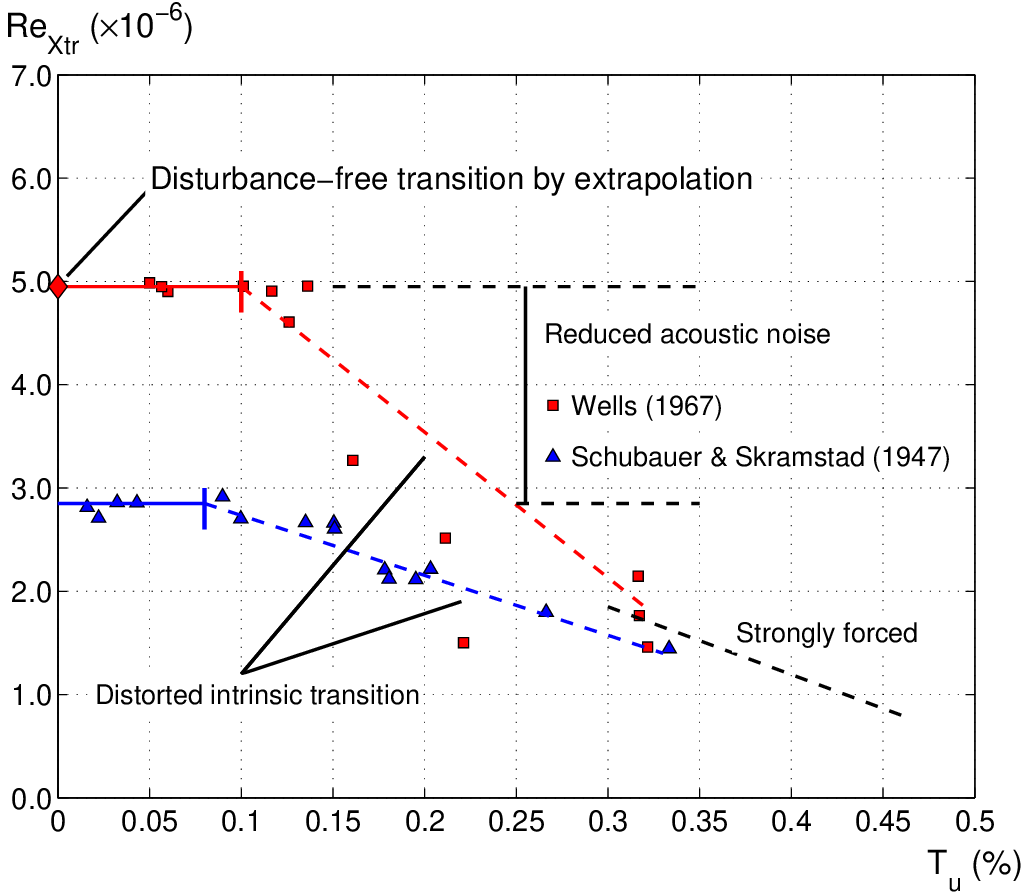}}
  \caption{Transition Reynolds number ($Re_{Xtr}$) as a function of free-stream turbulence level $T_{{\bfu}}$ for flat-plate boundary layers. In each case, the intrinsic turbulence transition is found to be independent of the threshold value $T^*_{{\bfu}}$ as shown by a small vertical line at $0.1\%$ and $0.08\%$. If the turbulence levels are substantially higher than $T^*_{{\bfu}}$, loss of laminar flow occurs prematurely but the actual process is mis-shaped and contorted by the test environment. Incoming flows with strong fluctuations can induce complete loss of laminar flow as the first two terms in our series solution (\ref{vort-sol}) contain convoluted vortex eddies of the free-stream turbulence. For the two perturbative environments commonly encountered in practice, disturbance-free transition is established at a {\it finite} value of $Re_{Xtr}$ (marked by a filled diamond) which is determined by extrapolating the test data to $T_{{\bfu}}=0$.} \label{xtr} 
\end{figure}

We draw our attention to a puzzling fact in flat-plate experiments: in order to observe the Tollmien-Schlichting waves, artificial excitations of specific frequencies had to be introduced into the laminar boundary layer so that the resulting measurements would agree with the predictions by linear stability theory which was an attempt to solve the complete Navier-Stokes equations {\it by approximation}. In the complete absence of the simulated excitations, instability waves could not be detected at the neighbourhood of the critical Reynolds number, as long as the external environmental disturbances were well-kept below a minimum, even though turbulence transition was unavoidable at some downstream location (Schubauer \& Skramstad 1947; Klebanoff {\it et al.} 1962). An explanation of this apparently paradoxical situation is that the Tollmien-Schlichting waves are usually too weak to measure with certainty (see, for instance, \S 17.7 of Tritton 1988). The question is: how can we distinguish the man-made disturbances from the ``genuine" instability waves? Evidently, if the flow has been disturbed, the measurement instrument ought to have picked up the responses of the laminar flow to the forced excitations since, in any event, the unstable waves are ineffective to impact.  
 
A comparable situation was realised in the leading-edge flow over a swept wing where the three-dimensional boundary layer was dominated by cross-flows (Deyhle \& Bippes 1996). In particular, controlled excitations or artificial disturbances had to be generated in order to observe cross-flow instabilities as stipulated in linear stability theory. The test conclusion says that free-stream turbulence affects transition indirectly, as long as $T_{{\bfu}} {<} 0.2\%$. It is also found that the receptivity effects due to acoustic excitations and non-uniformities of mean flow are weak. The cross-flow instabilities are seen to be less vulnerable to free-steam turbulence but the trend is in accord with the flat-plate experiments. By analogy, the main results may be interpreted as a confirmation that unstable cross-flow waves as predicted by the linear theory are unlikely to be observable in the low turbulence test environments. 

To obtain a solution from the equations of motion, any spatio-temporal flow modification, due to the presence of a disturbance during evolution, requires revised initial data, and a ``new" solution of the governing equations must be sought. This is particularly relevant to experiments where injections are used as a means of creating finite disturbances on pipe flow. Since the mass addition must be taken into account as $\nabla.\bfu {=} f(x){\neq}0$, any deviation from the continuity hypothesis must be assessed locally in a control volume analysis. The vorticity field is no longer solenoidal, and a modification in the no-slip boundary condition has to be considered. As a result of the problem reformulation, there are extra terms in the equations of vorticity dynamics which can be aggressive to reorganise the overall flow evolution. The forced phenomena observed in these related experimental works need to be quantified in dedicated studies.
\section{Conclusion}
We have shown that the flow evolution in a circular pipe from given initial state can be described by solution of the Navier-Stokes equations. The general solution of the vorticity equation offers a theoretical explanation to the experimental observations of Reynolds (1883). The non-linear effects specify a dynamic process of proliferating vorticity scales. We have also demonstrated that the vorticity production at pipe wall cannot be prescribed {\it a priori}, but can be determined by iteration. It is a matter of computations to map out the detailed flow structures. In essence, the vorticity solution not only explicates {\it turbulence}, but also clarifies the space-time gradation of the flow evolution. Specifically, the laminar-turbulent transition is understood to be an intrinsic part of the process, should the initial vorticity or the Reynolds number be sufficiently large. It is inevitable for laminar, transitional and turbulent motions to co-exist over a large part of the pipe, so that the flow field is normally intermittent in space and in time. In light of test evidence, and our analytical results, we assert that the intrinsic turbulence transition is primarily instigated by the non-linearity. Disturbances, either infinitesimal or finite, misrepresent the transition. On practical as well as theoretical grounds, it is important to understand the transition dynamics from the primitive equations, without recourse to modelling, because exact vorticity solutions can be obtained. Consequently, the impact of the non-linear effects on the flow development can be appreciated with confidence. Given the growing intricacy in flow field, prior to fully random turbulent state, experimental investigations on the transition process are surely overwhelmed by aberrant particulars, and hence have to, more or less, rely on statistical inference.
\vspace{8mm}
\begin{acknowledgements}
\noindent 
09 January 2019

\noindent 
\texttt{f.lam11@yahoo.com}
\end{acknowledgements}
\newpage
\appendix{Zeros for Bessel functions $J_n(\sigma_{n,k})$ and $J'_n(\alpha_{n,m})$} \label{app:a}
For every $n$, the zeros of the Bessel function $J_n(\sigma_{n,1})=0$ may be arranged in ascending order,
$0 < \sigma_{n,1} < \sigma_{n,2} < \sigma_{n,3} < \cdots  < \sigma_{n,k} < \cdots$. The ascending order also holds for the zeros of $J'_n(\alpha_{n,1})=0$.
As $n \rightarrow \infty$, the zeros can be estimated by the asymptotic formula (Olver {\it et al.} 2010)
\begin{equation*}
{\tilde\sigma}_{n,1} \; \sim \; n + 1.8557571 n^{1/3}+1.03315/n^{1/3}-0.00397/n,
\end{equation*}
and
\begin{equation*}
{\tilde\alpha}_{n,1} \; \sim \; n + 0.8086165 n^{1/3}+0.07249/n^{1/3}-0.05097/n.
\end{equation*}
We list the first few zeros of $J_n(\sigma_{n,1})=0$ and $J'_n(\alpha_{n,1})=0$ in the table below.
\vspace{5mm}

\begin{minipage}[h]{6cm}
\begin{center}
\begin{tabular}{ccc} \hline \hline
 $n$   &   $\sigma_{n,1}$ & $\;\;\;\;\;\;{\tilde\sigma}_{n,1}$\\ \hline \hline
     0    &         2.404826    &         -    \\ 
     1    &         3.831706    &         3.884937    \\ 
     2    &         5.135622    &         5.156134    \\ 
     3    &         6.380162    &         6.391488    \\ 
     4    &         7.588342    &         7.595682    \\ 
     5    &         8.771484    &         8.776696    \\ 
     6    &         9.936110    &         9.940037    \\ 
     7    &        11.086370    &        11.089456    \\ 
     8    &        12.225092    &        12.227593    \\ 
     9    &        13.354300    &        13.356376    \\ 
    10    &        14.475501    &        14.477256    \\ 
    11    &        15.589848    &        15.591356    \\ 
    12    &        16.698250    &        16.699562    \\ 
    13    &        17.801435    &        17.802589    \\ 
    14    &        18.899998    &        18.901023    \\ 
    15    &        19.994431    &        19.995348    \\ 
    16    &        21.085146    &        21.085973    \\ 
    17    &        22.172495    &        22.173244    \\ 
    18    &        23.256776    &        23.257459    \\ 
    19    &        24.338250    &        24.338876    \\ 
    20    &        25.417141    &        25.417717    \\ 
    21    &        26.493647    &        26.494180    \\ 
    22    &        27.567944    &        27.568438    \\ 
    23    &        28.640185    &        28.640644    \\ 
    24    &        29.710509    &        29.710937    \\ 
    25    &        30.779039    &        30.779440    \\ 
    26    &        31.845887    &        31.846264    \\ 
    27    &        32.911154    &        32.911508    \\ 
    28    &        33.974930    &        33.975264    \\ 
    29    &        35.037299    &        35.037614    \\ 
    30    &        36.098337    &        36.098635    \\ \hline \hline
\end{tabular}
\end{center}
\end{minipage}
\begin{minipage}[h]{6cm}
\begin{center}
\begin{tabular}{ccc} \hline \hline
 $n$   &   $\alpha_{n,1}$ & $\;\;\;\;\;\;{\tilde\alpha}_{n,1}$\\ \hline \hline
     0    &         0.000000    &         -    \\ 
     1    &         1.841184    &         1.830137    \\ 
     2    &         3.054237    &         3.050843    \\ 
     3    &         4.201189    &         4.199499    \\ 
     4    &         5.317553    &         5.316522    \\ 
     5    &         6.415616    &         6.414913    \\ 
     6    &         7.501266    &         7.500752    \\ 
     7    &         8.577836    &         8.577441    \\ 
     8    &         9.647422    &         9.647107    \\ 
     9    &        10.711434    &        10.711176    \\ 
    10    &        11.770877    &        11.770661    \\ 
    11    &        12.826491    &        12.826308    \\ 
    12    &        13.878843    &        13.878685    \\ 
    13    &        14.928374    &        14.928237    \\ 
    14    &        15.975439    &        15.975317    \\ 
    15    &        17.020323    &        17.020215    \\ 
    16    &        18.063265    &        18.063168    \\ 
    17    &        19.104462    &        19.104375    \\ 
    18    &        20.144083    &        20.144003    \\ 
    19    &        21.182270    &        21.182197    \\ 
    20    &        22.219146    &        22.219080    \\ 
    21    &        23.254821    &        23.254759    \\ 
    22    &        24.289386    &        24.289329    \\ 
    23    &        25.322924    &        25.322872    \\ 
    24    &        26.355510    &        26.355461    \\ 
    25    &        27.387207    &        27.387161    \\ 
    26    &        28.418075    &        28.418032    \\ 
    27    &        29.448165    &        29.448125    \\ 
    28    &        30.477526    &        30.477488    \\ 
    29    &        31.506199    &        31.506163    \\ 
    30    &        32.534224    &        32.534190    \\ \hline \hline
\end{tabular}
\end{center}
\end{minipage}
\newpage
%
%
\addcontentsline{toc}{section}{\noindent{References}}

\label{lastpage}
\end{document}